\documentclass[a4paper,11pt]{article}
\pdfoutput=1
\usepackage{jheppub}
\usepackage{graphicx,color,amsfonts,amsmath,amssymb}
\usepackage[normalem]{ulem}

\title{On the evolution process of two-component dark matter in the Sun
}

\author[a]{Chian-Shu Chen}
\author[b,1]{and Yen-Hsun Lin\note{Corresponding author.}}

\affiliation[a]{Department of Physics, Tamkang University,\\New Taipei 251, Taiwan}
\affiliation[b]{Department of Physics, National Cheng Kung University,\\Tainan 701, Taiwan}

\emailAdd{chianshu@gmail.com}
\emailAdd{yenhsun@phys.ncku.edu.tw}

\abstract{
We introduce dark matter (DM) evolution process in the Sun
under a two-component DM (2DM) scenario. Both DM species $\chi$ and $\xi$ with masses
heavier than $1\,{\rm GeV}$ are considered. In this picture, both species could be
captured by the Sun through DM-nucleus scattering and DM self-scatterings, 
e.g.~$\chi\chi$
and $\xi\xi$ collisions. In addition, the heterogeneous self-scattering due
to $\chi$ and $\xi$ collision is
essentially possible in any 2DM models. This new introduced scattering naturally
weaves the evolution processes of the two DM species that was assumed to evolve 
independently. Moreover, the heterogeneous self-scattering enhances the number of
DM being captured in the Sun mutually. This effect significantly exists in a broad 
range of DM mass spectrum. We have studied this phenomena and its implication for the
solar-captured DM annihilation rate. It would be crucial to the DM indirect detection
when the two masses are close.
General formalism of the 2DM evolution in the Sun as well as its kinematics are studied.}


\begin{document}
\maketitle
\flushbottom
\section{Introduction}

Dark matter (DM) composes five times as prevalent as ordinary matter,
yet its particle nature is still elusive. The essence of DM
is often portrayed as Weakly Interacting Massive Particle (WIMP)
with one specie. 
Experiments built to
probe the interaction between the Standard Model (SM) particles and DM are running in progress \cite{Aad:2015zva,Abdallah:2015ter,Akerib:2016vxi,Akerib:2017kat,Aprile:2017ngb,Amole:2017dex}. Besides, terrestrial
neutrino detectors \cite{Aartsen:2014oha,Aartsen:2016zhm,Choi:2015ara} and satellite detectors
\cite{Aguilar:2015ctt,TheFermi-LAT:2017vmf,Ambrosi:2017wek} are designed to detect the SM particle fluxes
from the annihilation of DM. Though much more stringent constraints
on the DM properties have been set, primary implication for DM is generally
from its gravitational influence. The understanding of DM is still
in its budding stage. 

Nevertheless, no strong evidence indicates that there exists only one-component DM  in the dark sector (DS). DM with \emph{n}-component (\emph{n}DM) is also a plausible
option.
Each one contributes relic abundance $\Omega_{\alpha}h^2$
to the the total relic abundance $\Omega_{\rm DM}h^2$ where $\Omega_{\rm DM}h^2=\sum_{\alpha}\Omega_{\alpha}h^2 \approx 0.12$ \cite{Ade:2013zuv} and $h$ is the Hubble constant.
Theoretical models on the
two-component DM (2DM) scenario have been proposed recently~
\cite{Hambye:2008bq,Lebedev:2011iq,Agashe:2014yua,Gross:2015cwa,Arcadi:2016kmk,Aoki:2016glu}. Works regard the 2DM models on the direct and indirect searches can be also found in refs.~\cite{Alhazmi:2016qcs,Bhattacharya:2016tma,Herrero-Garcia:2017vrl} and references therein. In addition, cosmological \emph{N}-body simulation incorporates
2DM that leads to the large scale structure which agrees the observation
has been done recently \cite{Medvedev:2013vsa}.

In this work, we study the DM evolution in the Sun under
the 2DM scenario with particle species $\chi$ and $\xi$. Recent studies
on solar captured 2DM are only a few, and, to our understanding, such
studies consider the evolution of two species that are independent
in the Sun. On the other hand, the collision between $\chi$ and $\xi$
does not account for the DM capture in the Sun. The collisions among
DM particles 
are generally characterized as DM self-interactions. In the 
1DM 
scenario, 
DM self-interaction \cite{Spergel:1999mh,Massey:2015dkw,Kahlhoefer:2015vua,Buckley:2009in,Aarssen:2012fx,Tulin:2012wi}
is addressed to alleviate the discrepancy between the collisionless
\emph{N}-body simulation and the observations. Such
inconsistency arises from  small-scale structure 
\cite{Navarro:1996gj,Moore:1994yx,Flores:1994gz,Randall:2007ph,Feng:2009hw,Walker:2011zu,Walker:2012td,BoylanKolchin:2011de,BoylanKolchin:2011dk,Elbert:2014bma}
could be resolved
by imposing the constraint $0.1<\sigma_{\rm DM}/m_{\rm DM}<10\,{\rm cm}^{2}\,{\rm g}^{-1}$,
where $\sigma_{\rm DM}$ and $m_{\rm DM}$ are DM self-interacting cross
section and mass respectively. In addition, 
by incorporating baryonic effect, that the problem of diverse galactic
rotation curves \cite{Oman:2015xda,Elbert:2016dbb} could be mitigated with the constraint $3<\sigma_{\rm DM}/m_{\rm DM}<6\,{\rm cm}^{2}\,{\rm g}^{-1}$ \cite{Kamada:2016euw,Creasey:2016jaq,Valli:2017ktb,Robertson:2017mgj}. 

The study of DM captured by the Sun has been 
investigated in
refs.~\cite{Griest:1986yu,Gould:1987ju,Gould:1987ir,Bernal:2012qh,Baum:2016oow,Fornengo:2017lax}. Updated calculation including the non-zero momentum
transfer and implication for DM-electron capture are also indicated in
ref.~\cite{Garani:2017jcj}.
Our work is based on the earlier ones, and we further extend the framework to the 2DM scenario.
The contribution to the capture rate from DM 
self-interaction denotes the self-capture in general. Furthermore, in the 2DM scenario,
a plausible situation is that 
aside from $\chi\chi$ and $\xi\xi$ collisions, the self-interactions can also happen between $\chi$ and $\xi$, the
\emph{heterogeneous} self-interaction. In this case, $\chi\xi$ collision
plays a role of heterogeneous self-capture. In the presence of heterogeneous
self-interaction, extra coupling terms should be incorporated in the
evolution equations. The evolution processes of $\chi$ and of $\xi$ cannot be treated separately.
In our analysis, we found that the number of the sub-dominant DM specie is subject to a correction
from the dominant DM specie.
Nevertheless, the heterogeneous self-interaction not only increases
the capture rate, it also responsible for the self-evaporation and
self-ejection effects.

The general formalism to calculate the 2DM evolution in the Sun is
given in this paper. Full expressions for the rates of DM-nucleus
capture, evaporation and contributions from DM self-interaction including
heterogeneous effects are presented. Although the framework
shown here focuses on the 2DM scenario, it can be easily generalized
to any \emph{n}DM scenario. Unless subtle interaction is specified
in the \emph{n}DM case, e.g.~3-body scattering, the results provided 
in this paper 
are generally applied to any 2-body scattering
with different masses. As a remark, when the 2DM scenario is invoked,
co-annihilation could happen if the masses are nearly degenerate between
$\chi$  and $\xi$ \cite{Binetruy:1983jf,Griest:1990kh,Bertone:2004pz}. However, in the later analysis, we scan a broad range
of DM mass spectrum. In most of the situation, the masses are not
degenerate. Thus, the co-annihilation can be considered irrelevantly.
Omitting it from our discussion is reasonable.
Works in regard to co-annihilation and its implication for the
solar-captured DM can be found in refs.~\cite{Blennow:2015xha,Blennow:2015hzp}.

This paper is structured as follows. In section~\ref{sec:2DM_remarks}, we briefly introduce 
the 2DM scenario 
including notations and general
assumptions in this work. In section~\ref{sec:formalism}, the evolution
equations for $\chi$ and $\xi$ are given. Coupling terms from heterogeneous
self-interaction are introduced in the equations. In section~\ref{sec:scattering_rates},
we present all the rates of interaction. Physical implications
are shown. In section~\ref{sec:numerical_analysis}, numerical results
of the evolution equations are calculated along with the DM total
annihilation rates for both species. If DM can annihilate to final state with SM particles,
the DM total annihilation rate characterizes the intensity of such
SM particle flux. Finally, we summarize in section~\ref{sec:Summary}.
Mathematical derivations of the rates relative to the heterogeneous
self-interaction are given in the appendix.

\section{Remarks on the 2DM scenario}\label{sec:2DM_remarks}
\subsection{Brief review of the 2DM models}
The 2DM models typically require two different discrete symmetries assigned to each DM to sustain their stability. However, simple extension of the SM group by a global discrete symmetry can be violated by gravity~\cite{Krauss:1988zc} or induce the cosmic defects that are not compatible with cosmological observations~\cite{Hindmarsh:1994re}. These problems can be evaded if one retains the wanted discrete symmetries by breaking a gauge group.  

One of the simplest 2DM models is based on an Abelian $U(1)_d$, where $d$ refers to hidden charge, and then assign some integer quantum numbers $n_{1}$ and $n_{2}$ to the hidden scalar fields $\chi$ and $\xi$ \cite{Batell:2010bp}. After $\chi$ and $\xi$ fields develop vacuum expectation values, the $U(1)_d$ will break with the residual discrete symmetries $Z_{n_{1}}\otimes Z_{n_{2}}$. Examples are $Z_{2}\otimes Z'_{2}$~\cite{Agashe:2014yua} or $Z_{2}\otimes Z_{4}$~\cite{Aoki:2016glu}. As a result $\chi$ and $\xi$ can be both stable and will be the DM candidates. Other interesting models are the DM can be the multiplet of vector bosons of some hidden non-Abelian gauge groups~\cite{Hambye:2008bq,Lebedev:2011iq,Gross:2015cwa,Arcadi:2016kmk}. For example, a hidden $SU(2)_d$ gauge group with a hidden fundamental representation of scalar field $\phi$, 
\begin{eqnarray}
\mathcal{L}_{d} = -\frac{1}{4}F'^{a\mu\nu}F'^a_{\mu\nu} + (D_{\mu}\phi)^{\dag}(D^{\mu}\phi) + V(\phi), 
\end{eqnarray}
where $a=1,2,3$, $F'^{a\mu\nu}$ is the hidden field strength, $D^{\mu}\phi = \partial^{\mu}\phi - i\frac{g_d}{2}\tau^a\cdot A'^{a\mu}\phi$, $g_d$ is the hidden coupling constant and $A'^{\mu}$ is the hidden gauge field. After $\phi$ developing a vacuum expectation value, $SU(2)_d$ breaks and the corresponding hidden gauge bosons will become degenerate massive particles. In such case, a residual custodial $SO(3)$ symmetry remains due to the fact of scalar field $\phi$ being the fundamental representation. A $Z_{2}\otimes Z'_{2}$ discrete symmetry, 
\begin{eqnarray}
Z_{2} &:& A'^{1}_{\mu} \rightarrow -A'^{1}_{\mu} \quad , \quad A'^{2}_{\mu} \rightarrow -A'^{2}_{\mu} ,\nonumber \\
Z'_{2}&:& A'^{1}_{\mu} \rightarrow -A'^{1}_{\mu} \quad , \quad A'^{3}_{\mu} \rightarrow -A'^{3}_{\mu}, 
\end{eqnarray}
will apply to the hidden gauge bosons. Therefore, $A'^{a}_{\mu}$ can be stable and DM candidates. One can extend the non-Abelian $SU(2)_d$ to larger groups~\cite{Gross:2015cwa}. It is worth of mentioning that in such models the DM can interact with the SM sectors via Higgs portal,
\begin{eqnarray}
\mathcal{L}_{\rm{Higgs}} \supset -\mu^2H^{\dag}H + \lambda(H^{\dag}H)^2 - \mu^2_{\phi}\phi^{\dag}\phi + \lambda_{\phi}(\phi^{\dag}\phi)^2 + \lambda_{H\phi}\phi^{\dag}\phi H^{\dag}H . 
\end{eqnarray}
or gauge boson kinetic mixing, 
\begin{eqnarray}
\mathcal{L}_{\rm{gauge}} \supset \varepsilon B_{\mu\nu}X^{\mu\nu} 
\end{eqnarray}
where $B_{\mu\nu}$ and $X_{\mu\nu}$ are the field strength of $U(1)_{Y}$ and $U(1)_d$ respectively. Therefore the DM annihilation final states are SM particles or new scalars if it is kinematically allowed. These DM particles behave as thermal WIMPs and can retain the observed DM relic abundance. Interesting
phenomena provided by these DM particles, e.g.~excess cosmic rays, direct search, cosmology and collider physics, can be found in refs.~\cite{Hambye:2008bq,Lebedev:2011iq,Mambrini:2011dw,Baek:2012se,Arina:2009uq,Ko:2014loa,Gross:2015cwa,Bernal:2015ova,Arcadi:2016kmk,Duch:2017khv,Heikinheimo:2018duk} and references therein.

\subsection{Notation conventions and general assumptions} 

Suppose the two DM species $\chi$ and $\xi$ only differ in
mass such that $m_{\chi}\neq m_{\xi}$ in general. The corresponding relic
abundances are $\Omega_{\alpha}h^2$ where $\alpha=\chi$ and $\xi$. Assuming
DM as the thermal relic and its total abundance $\Omega_{\rm DM}h^2$ is made
up of $\Omega_{\chi}h^2$ and $\Omega_{\xi}h^2$, e.g.~$\Omega_{\rm DM}h^2=\Omega_{\chi}h^2+\Omega_{\xi}h^2$.
Ratio of the two relic abundances can be defined by
\begin{equation}
r_{\rho}=\frac{\Omega_{\xi}}{\Omega_{\chi}}.
\end{equation}
If the annihilation is dominated by $s$-wave process at freeze-out
epoch, we have the relic abundance inversely proportional to its thermal
relic annihilation cross section $\langle\sigma v\rangle_{0}$ that
is given by \cite{Griest:1986yu}
\begin{equation}
\Omega_{\alpha}h^2\propto\frac{1}{\langle\sigma_{\alpha}v\rangle_{0}}\label{eq:Omega_a}
\end{equation}
which is mass-independent up to logarithmic corrections. Hence, we
can express the annihilation cross sections by $r_{\rho}$:
\begin{equation}
r_{\rho}=\frac{\langle\sigma_{\chi}v\rangle_{0}}{\langle\sigma_{\xi}v\rangle_{0}}.\label{eq:r_sigv}
\end{equation}
Therefore, with eqs.~(\ref{eq:Omega_a}) and (\ref{eq:r_sigv}),
we can further define an \emph{effective} annihilation cross section $\langle\sigma_{{\rm eff}}v\rangle_{0}$
for $\Omega_{\rm DM}h^2$ by
\begin{equation}
\Omega_{\rm DM}h^2=\Omega_{\chi}h^2+\Omega_{\xi}h^2\propto\frac{1+r_{\rho}}{\langle\sigma_{\chi}v\rangle_{0}}\equiv\frac{1}{\langle\sigma_{{\rm eff}}v\rangle_{0}}.\label{eq:freeze_sigv}
\end{equation}
To produce thermal relic abundance $\Omega_{\rm DM}h^2\approx0.12$ \cite{Ade:2013zuv},
it is reasonable to assume $\langle\sigma_{{\rm eff}}v\rangle_{0}\approx3\times10^{-26}\,{\rm cm}^{3}\,{\rm s}^{-1}.$

\begin{figure}
\begin{centering}
\includegraphics[width=0.45\textwidth]{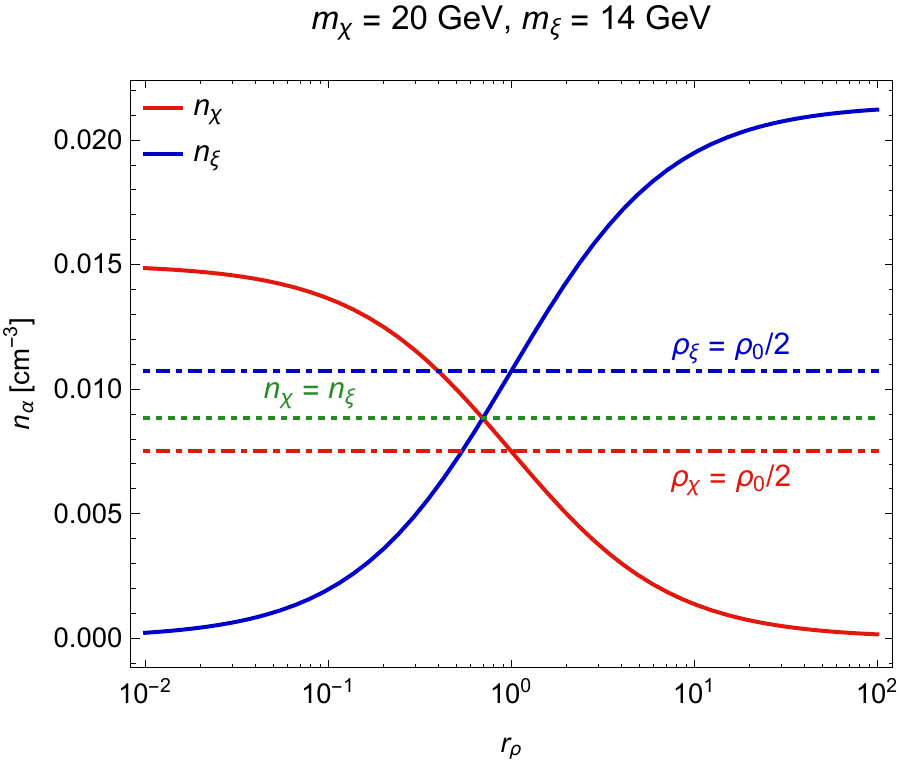}\quad\includegraphics[width=0.45\textwidth]{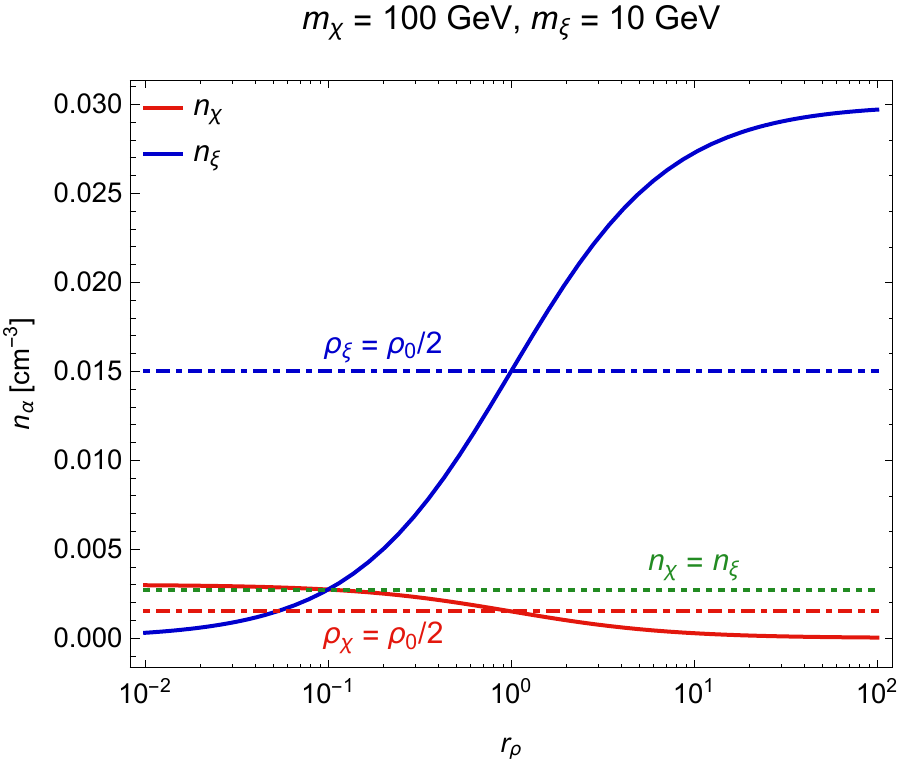}
\par\end{centering}
\caption{\label{fig:number_dens}Local DM number density $n_{\alpha}$ vs.~$r_{\rho}$.
$\rho_{\rm DM}=0.3\,{\rm GeV}\,{\rm cm}^{-3}$ is the local DM density
near the solar neighborhood. A similar result can be also found in the figure 1 of ref.~\cite{Herrero-Garcia:2017vrl}.}

\end{figure}

In addition, the number of DM particles captured by the Sun is relevant
to the local  DM density $\rho_{\rm DM}$ around our solar neighborhood.
Thus, the relation 
\begin{equation}
\rho_{\rm DM}=0.3\,{\rm GeV}\,{\rm cm}^{-3}=\rho_{\chi}+\rho_{\xi}
\end{equation}
that should hold. Without loss of generality, we let $\rho_{\alpha}\propto\Omega_{\alpha}h^2$,
thus $r_{\rho}=\rho_{\xi}/\rho_{\chi}$. We can rewrite the above identity
as
\begin{equation}
\rho_{\rm DM}=\rho_{\chi}(1+r_{\rho})=0.3\,\rm{GeV}\,\rm{cm}^{-3}\label{eq:rho_x}
\end{equation}
and the local DM number density $n_{\alpha}=\rho_{\alpha}/m_{\alpha}$
is plotted in figure~\ref{fig:number_dens} versus $r_{\rho}$. A similar plot can
be found in the figure 1 of ref.~\cite{Herrero-Garcia:2017vrl} as well. When
$r_{\rho}=1$, the DM with lighter mass has the higher number density.
On the other hand, the number densities happen to be equal when $r_{\rho}=m_{\xi}/m_{\chi}$.
In the later analysis, we 
assume the validations of
eqs.~(\ref{eq:r_sigv}), (\ref{eq:freeze_sigv}) and (\ref{eq:rho_x})
pass to the present day. Once $r_{\rho}$ is assigned, $\langle\sigma_{\alpha}v\rangle$
and $\rho_{\alpha}$ are specified consequently.

\section{General formalism of dark matter evolution in the Sun\label{sec:formalism}}

\subsection{The 1DM evolution equation
}

When the Sun sweeps the DM halo, DM particles are attracted by the
solar gravity. The subsequent scatterings with the solar nuclei and
other DM particles already trapped in the Sun could happen. DM particles can
be captured by the Sun when its final velocity is smaller than the
escape velocity of the Sun after scattering. Alternatively, DM particles
trapped inside the Sun will be kicked out if its final velocity
after the scattering with the nuclei is larger than the escape velocity.
The inclusion of DM self-interaction will also have effects on the
capture and evaporation of DM particles in the Sun. Incorporating all these
effects, the general equation describes the DM evolution process is given by
\begin{equation}
\frac{dN_{\rm DM}}{dt}=C_{c}+(C_{s}-C_{e})N_{\rm DM}-(C_{a}+C_{se})N_{\rm DM}^{2}\label{eq:evo_eq}
\end{equation}
where $N_{\rm DM}$ is the DM number in the Sun, $C_{c}$ the rate at
which DM is captured by the solar nuclei \cite{Gould:1987ju,Gould:1987ir,Bernal:2012qh,Garani:2017jcj},
$C_{s}$ the self-capture rate at which DM is captured due to scattering
with other trapped DM inside the Sun \cite{Zentner:2009is}, $C_{e}$
the evaporation rate due to DM-nucleus scattering \cite{Gould:1989tu},
$C_{a}$ the annihilation and $C_{se}$ the self-evaporation rate
that caused by DM-DM scattering \cite{Chen:2014oaa}.

However, recent study shows unless DM mass $m_{\rm DM}\lesssim4\,{\rm GeV}$,
the evaporation effect is much inefficient even with the inclusion
of $C_{se}$ \cite{Chen:2014oaa}. Thus, in the absence of evaporations,
eq.~(\ref{eq:evo_eq}) reads

\begin{equation}
\frac{dN_{\rm DM}}{dt}=C_{c}+C_{s}N_{\rm DM}-C_{a}N_{\rm DM}^{2}\label{eq:evo_no_eva}
\end{equation}
along with an analytical solution
\begin{equation}
N_{\rm DM}=\frac{C_{c}\tanh(t/\tau)}{\tau^{-1}-C_{s}\tanh(t/\tau)/2},\label{eq:1DM_N}
\end{equation}
where $\tau=1/\sqrt{C_{c}C_{a}+C_{s}^{2}/4}$ is the equilibrium timescale.
When $t\gg\tau$, $dN_{\rm DM}/dt=0$.

\subsection{The 2DM evolution equations
}

On the other hand, eq.~(\ref{eq:evo_eq}) only characterizes 1DM scenario. Once the second DM specie is included, additional evolution equation
should be added. In this scenario, the self-interactions are not only
due to $\chi\chi$ and $\xi\xi$ scatterings as well as $\chi\xi$ scattering. Thus, the evolution
process is determined by
\begin{subequations}
\begin{align}
\frac{dN_{\chi}}{dt}&=  C_{c}^{\chi}+(C_{s}^{\chi}-C_{e}^{\chi})N_{\chi}
  +(C_{s}^{\chi\to \xi}-C_{se}^{\chi\to \xi}N_{\chi})N_{\xi}-(C_{a}^{\chi}+C_{se}^{\chi})N_{\chi}^{2},\label{eq:A_evo}\\
\frac{dN_{\xi}}{dt}&=  C_{c}^{\xi}+(C_{s}^{\xi}-C_{e}^{\xi})N_{\xi}
+(C_{s}^{\xi\to \chi}-C_{se}^{\xi\to \chi}N_{\xi})N_{\chi}-(C_{a}^{\xi}+C_{se}^{\xi})N_{\xi}^{2}.\label{eq:B_evo}
\end{align}
\end{subequations}
The above equations are modified from eq.~(\ref{eq:evo_eq}).
Four additional coefficients are introduced. $C_{s}^{\chi(\xi)\to \xi(\chi)}$
denote the heterogeneous self-capture rates due to halo $\chi(\xi)$ scatters with trapped
$\xi(\chi)$ in the Sun. $C_{se}^{\chi(\xi)\to \xi(\chi)}$ denote the heterogeneous self-evaporation
rates due to the $\chi(\xi)$ scatters with $\xi(\chi)$ in the Sun. The rest
are $C_{c}^{\alpha}$ the solar captures, $C_{s}^{\alpha}$ the self-capture
rates, 
$C_{e}^{\alpha}$ the evaporation
rates, $C_{se}^{\alpha}$ the self-evaporation rates and $C_{a}^{\alpha}$ the annihilation rates as those in the 1DM case. 

Both eqs.~(\ref{eq:A_evo}) and (\ref{eq:B_evo}) correlate together
through the terms subject to 
$N_\xi$ in $dN_\chi/dt$ and $N_\chi$ in $dN_\xi/dt$.
The DM numbers of $\chi$
and of $\xi$ in the Sun are mutually dependent. Without correlation
terms, the evolution processes for both DM species are decoupled.
Generally, evaporation is inefficient unless the DM mass is light
enough, typically when $m_{\alpha}^{{\rm ev}}\lesssim4\,{\rm GeV}$.
Even including the extra contribution $C_{se}^{\chi(\xi)\to \xi(\chi)}$, we
have numerically justified that this effect does not change the $m_{\alpha}^{{\rm ev}}$
dramatically. Thus, when $m_{\alpha}>4\,{\rm GeV}$, we can safely
ignore the evaporation from eqs.~(\ref{eq:A_evo}) and (\ref{eq:B_evo}).
Therefore,
\begin{subequations}
\begin{align}
\frac{dN_{\chi}}{dt}&=C_{c}^{\chi}+C_{s}^{\chi}N_{\chi}+C_{s}^{\chi\to \xi}N_{\xi}-C_{a}^{\chi}N_{\chi}^{2},\label{eq:A_no_eva}\\
\frac{dN_{\xi}}{dt}&=C_{c}^{\xi}+C_{s}^{\xi}N_{\xi}+C_{s}^{\xi\to \chi}N_{\chi}-C_{a}^{\xi}N_{\xi}^{2}.\label{eq:B_no_eva}
\end{align}
\end{subequations}
But in the later numerical
calculations, we will always use the general expressions eqs.~(\ref{eq:A_evo}) and (\ref{eq:B_evo}).  Note that the evolution equations have no analytical expressions
on $N_{\alpha}$ and $\tau$ in the 2DM scenario. However, approximated expressions can be obtained in certain situations. It will be discussed
in section~\ref{sec:numerical_analysis}.


\section{Dark matter scattering rates\label{sec:scattering_rates}}

\subsection{The solar capture rate}

The solar capture rate due to DM-nucleus scattering can be numerically
approximated as \cite{Jungman:1995df,Bertone:2004pz}
\begin{equation}
C_{c}^{{\rm SI}}\simeq4.1\times10^{24}\,{\rm s}^{-1}\,\left(\frac{\rho_{\alpha}}{{\rm GeV}\,{\rm cm}}\right)\left(\frac{270\,{\rm km}\,{\rm s}^{-1}}{\bar{v}}\right)^{3}\left(\frac{\sigma_{{\rm H}}^{{\rm SI}}+0.175\sigma_{{\rm He}}^{{\rm SI}}}{10^{-6}\,{\rm pb}}\right)\left(\frac{{\rm GeV}}{m_{\alpha}}\right)^{2}
\end{equation}
for spin-independent (SI) case and
\begin{equation}
C_{c}^{{\rm SD}}\simeq1.12\times10^{25}\,{\rm s}^{-1}\,\left(\frac{\rho_{\alpha}}{{\rm GeV}\,{\rm cm}}\right)\left(\frac{270\,{\rm km}\,{\rm s}^{-1}}{\bar{v}}\right)^{3}\left(\frac{\sigma_{{\rm H}}^{{\rm SD}}}{10^{-6}\,{\rm pb}}\right)\left(\frac{{\rm GeV}}{m_{\alpha}}\right)^{2}
\end{equation}
for spin-dependent (SD) case. $\rho_{\alpha}$ is the DM local density
and $\bar{v}=270\,{\rm km}\,{\rm s}^{-1}$ the DM velocity dispersion.
$\sigma_{{\rm H, He}}^{{\rm SI,SD}}$ is the DM-nucleus scattering cross section for hydrogen or helium. Taking proton mass $m_p$ is close to neutron mass
$m_n$.
The DM-nucleus cross section $\sigma_{A}$ at which interaction
is undergoing that is related to DM-nucleon cross section $\sigma_{\alpha p}$
by
\begin{equation}
\sigma_{A}^{{\rm SI}}=A^{2}\left(\frac{m_{A}}{m_{p}}\right)^{2}\left(\frac{m_{\alpha}+m_{p}}{m_{\alpha}+m_{A}}\right)^{2}\sigma_{\alpha p}^{{\rm SI}}
\end{equation}
for SI interaction interaction and
\begin{equation}
\sigma_{A}^{{\rm SD}}=A^{2}\left(\frac{m_{\alpha}+m_{p}}{m_{\alpha}+m_{A}}\right)^{2}\frac{4(J+1)}{3J}|\langle S_{p}\rangle+\langle S_{n}\rangle|^{2}\sigma_{\alpha p}^{{\rm SD}}
\end{equation}
for SD interaction, where $A$ is the atomic number, $m_{A}$ the corresponding nucleus mass, $J$ the total
angular momentum of the nucleus and $\langle S_{p}\rangle$ and $\langle S_{n}\rangle$
the spin expectation values of proton and of neutron averaged over
the entire nucleus \cite{Engel:1989ix,Ellis:1987sh,Pacheco:1989jz,Engel:1992bf,Divari:2000dc,Bednyakov:2004xq}.
To apply the above results, we have assumed $\chi$ and $\xi$ obey the
same Maxwell-Boltzman velocity distribution. The effect of uncertainties in
velocity distributions to the capture rate is minor \cite{Choi:2013eda}.
We note that
$\sigma_{\alpha p}$ is a model-dependent parameter in general.

In addition, following earlier work \cite{Jungman:1995df}, refined
calculation on solar capture rate including the contributions from
elements beyond hydrogen and helium can be found in refs.~\cite{Bernal:2012qh,Garani:2017jcj}.
In ref.~\cite{Garani:2017jcj}, constant scattering cross section
as well as velocity-dependent and transfer-momentum-dependent cases
are fully considered. We adopt the numerical procedure in ref.~\cite{Garani:2017jcj} to calculate $C_c$ in the 2DM scenario.

\subsection{The self-capture rate\label{subsec:DM_self_cap}}

The self-capture happens when the halo DM particles scatter off the DM particles that are already trapped inside the Sun.
Starting
with halo $\chi$ particle captured by the trapped $\xi$ particle in
the Sun. Therefore, the coefficient of heterogeneous self-capture rate
can be expressed as
\begin{equation}
C_{s}^{\chi\to \xi}=\frac{\int 4\pi r^{2}(dC_{s}^{\chi\to \xi}/dV)dr}{\int 4\pi r^{2}n_{\xi}(r)dr}\label{eq:Cs_cap}
\end{equation}
where $dC_{s}^{\chi\to \xi}/dV$ is the heterogeneous self-capture rate in the Sun
within a given shell. It is determined by $m_\chi$, $m_\xi$ and $\sigma_{\chi\xi}$ where $\sigma_{\chi\xi}$ is the heterogeneous self-scattering cross section. The analytical form of $dC_{s}^{\chi\to \xi}/dV$
is given in eq.~(\ref{eq:dCs/dV_general}).
Assuming the heat exchanges among DM particles are very efficient after
capture. They will quickly reach the thermal equilibrium temperature $T_{\alpha}=T$.
Therefore, $n_{\alpha}(r)=n_{\alpha}^{0}e^{-m_{\alpha}\phi(r)/T}$
where $n_{\alpha}^{0}$ is the DM number density in the solar core,
$\phi(r)=\int_{0}^{r}GM(r^{\prime})/r^{\prime2}dr^{\prime}$ and $M(r^{\prime})$
the solar mass enclosed by radius $r^{\prime}$.
The case for halo $\xi$
particle captured by trapped $\chi$ particle is essentially the same.
Simply swaps $\chi$ and $\xi$ and replace all $\chi$'s parameters
by $\xi$'s.

\begin{figure}
\begin{centering}
\includegraphics[width=0.5\textwidth]{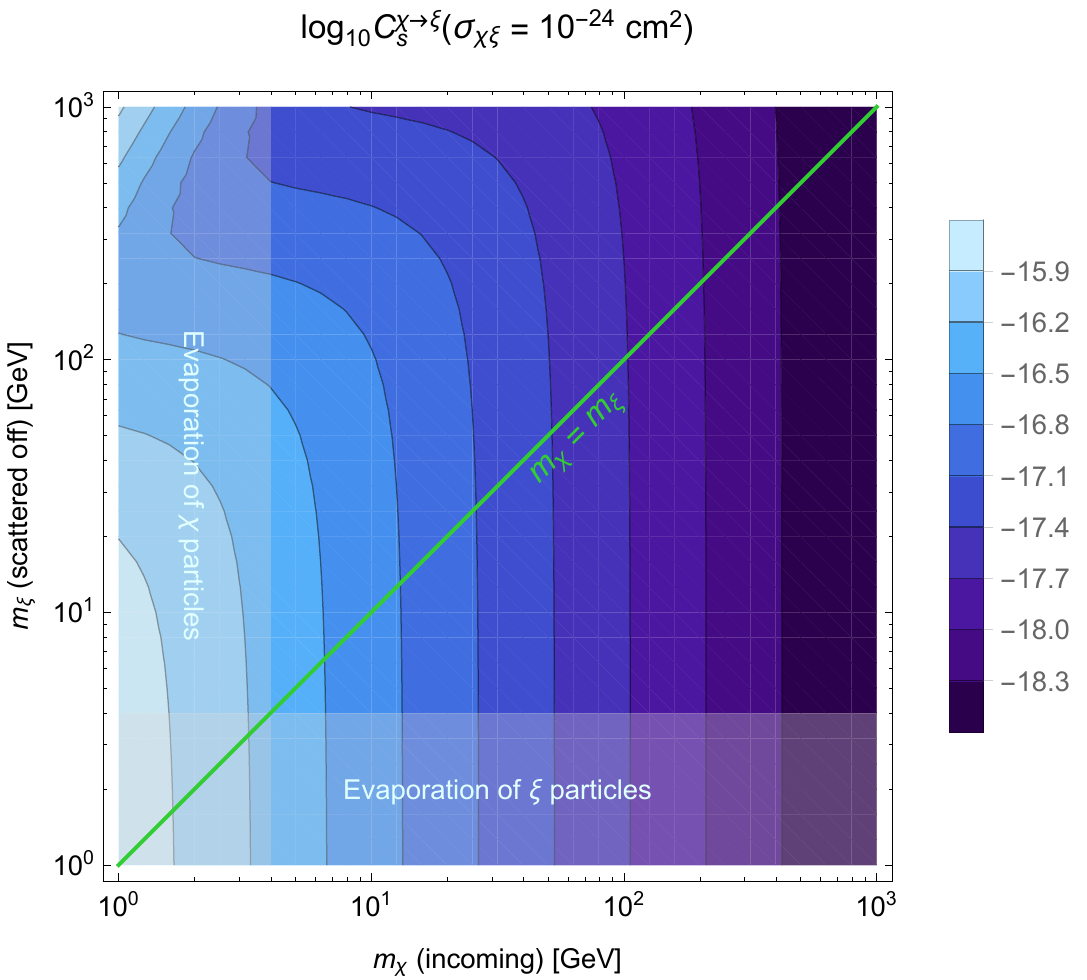}
\par\end{centering}
\caption{\label{fig:Cs}Self-capture coefficient $C_{s}^{\chi\to \xi}$ and $\sigma_{\chi\xi}=10^{-24}\,{\rm cm}^{2}$.
The green line denotes the self-capture as a result of the same DM
specie. White shaded region is the evaporation dominant region. Null capture
happens here.}
\end{figure}

The expression given in eq.~(\ref{eq:Cs_cap}) is generally for $m_{\chi}\neq m_{\xi}$.
Nevertheless, if the capture is due to the same DM specie, e.g.~$\chi\chi$
or $\xi\xi$ scatterings, it is done by letting $m_{\chi}=m_{\xi}$ and denotes
as $C_{s}^{\alpha}$. Such that we have a rather simple analytical
expression \cite{Zentner:2009is}:
\begin{gather}
C_{s}^{\alpha}=\sqrt{\frac{3}{2}}n_{\alpha}\sigma_{\alpha}v_{{\rm esc}}(R_{\odot})\frac{v_{{\rm esc}}(R_{\odot})}{\bar{v}}\langle\hat{\phi}_{\alpha}\rangle\frac{{\rm erf}(\eta)}{\eta}
\end{gather}
where $\sigma_{\alpha}$ is the self-scattering cross
section, $\langle\hat{\phi}_{\alpha}\rangle\simeq5.1$ \cite{Gould:1987ju}
a dimensionless average solar potential experienced by the captured
DM within the Sun and $v_{{\rm esc}}(R_{\odot})\approx632\,{\rm km}\,{\rm s}^{-1}$
the Sun's escape velocity at surface. We will characterize $\sigma_{\alpha}$
by
\begin{equation}
\sigma_{\alpha}/m_{\alpha}\approx3\,{\rm cm}^{2}\,{\rm g}^{-1}\label{eq:sig_T}
\end{equation}
in the later numerical analysis. Such relation appears to alleviate
the diversity of galactic rotation curve in the presence of baryonic
effect as well as small-scale structure problems \cite{Kamada:2016euw}.
Plot for $C_{s}^{\chi\to \xi}$ in the $m_{\chi}-m_{\xi}$ plane is shown in figure~\ref{fig:Cs}.

\subsection{The annihilation rate}

When more and more DM particles accumulate in the Sun,
the rate of annihilation becomes stronger.
The coefficient of annihilation rate is expressed as
\begin{equation}
C_{a}^{\alpha}=\langle\sigma_{\alpha}v\rangle\frac{\int_{0}^{R_{\odot}}n_{{\rm \alpha}}^{2}(r)4\pi r^{2}dr}{[\int_{0}^{R_{\odot}}n_{\alpha}(r)4\pi r^{2}dr]^{2}}
\end{equation}
where $\langle\sigma v\rangle$ is the DM annihilation cross section.
An approximation
for $C_{a}^{\alpha}$ is given by \cite{Griest:1986yu} \begin{subequations}
\begin{equation}
C_{a}^{\alpha}=\langle\sigma_{\alpha}v\rangle\frac{V_{2}}{V_{1}^{2}}
\end{equation}
where
\begin{equation}
V_{j}\approx6.8\times10^{28}\,{\rm cm}^{3}\,\left(\frac{T}{T_{\odot}}\right)^{3/2}\left(\frac{10\,{\rm GeV}}{jm_{\alpha}}\right)^{3/2},\quad j=1,2;
\end{equation}
\end{subequations}is the DM effective volume and $T_{\odot}=1.54\times10^{7}\,{\rm K}$
the solar core temperature. Essentially, the DM temperature $T$ is not necessary
the same as the solar temperature $T_{\odot}$ ~\cite{Garani:2017jcj,Chen:2015poa}
and depends on $m_{\alpha}$. 
But the temperature deviation from $T_\odot$ is
generally small and has little impact on the final number of DM 
particles in the Sun. It is reasonable to impose $T=T_\odot$ in our
later discussion.

\subsection{The evaporation and self-ejection rates}

In the Sun, after the collision happens between two particles, if one gets velocity
larger than the escape velocity $v_{{\rm esc}}$, it won't be captured. 
This effect is called evaporation. When the evaporation is caused by DM-nucleus scattering, an approximation
is given in ref.~\cite{Busoni:2013kaa}
\begin{equation}
C_{e}^{\alpha}\simeq\frac{8}{\pi^{2}}\sqrt{\frac{2m_{\alpha}}{\pi T}}\frac{v_{{\rm esc}}^{2}(0)}{\bar{r}^{3}}e^{-m_{\alpha}v_{{\rm esc}}^{2}(0)/2T}\Sigma_{{\rm evap}},\label{eq:Ce}
\end{equation}
where $v_{{\rm esc}}(0)=1366\,{\rm km}\,{\rm s}^{-1}$ is the escape
velocity at solar core, $\bar{r}$ the mean DM distance from the solar
center. The quantity $\Sigma_{{\rm evap}}$ is a factor that relates
to the DM-nucleus scattering cross section \cite{Busoni:2013kaa}.
The expression $C_{e}^{\alpha}$ is valid when $m_{\alpha}/m_{A}>1$.
In our calculation, we modified the numerical procedure done in ref.~\cite{Garani:2017jcj}
for the 2DM scenario. 

\begin{figure}
\begin{centering}
\includegraphics[width=0.45\textwidth]{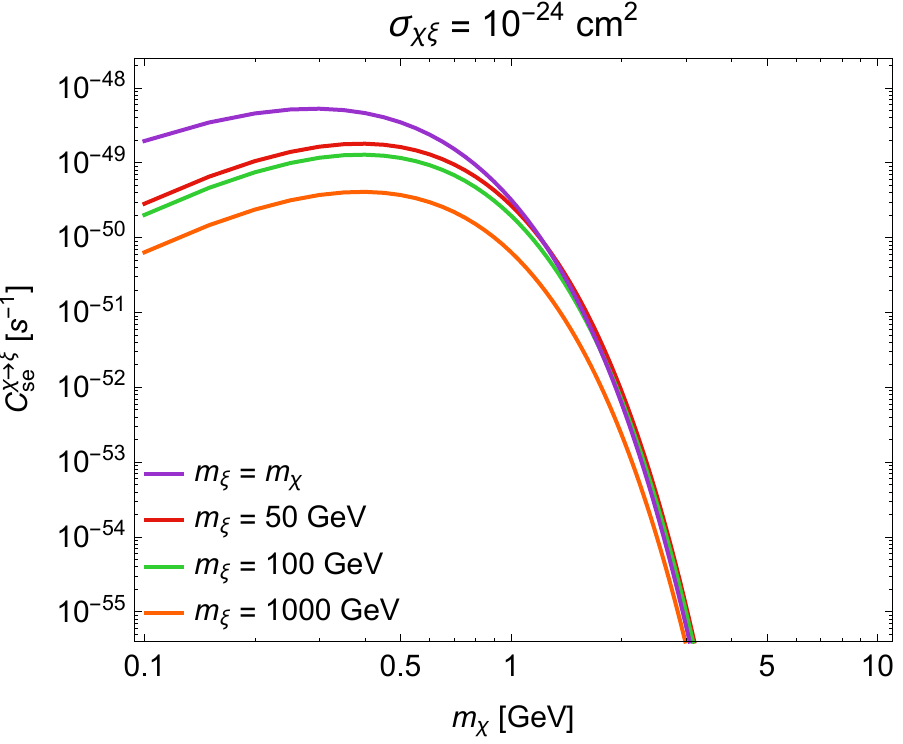}
\par\end{centering}
\caption{\label{fig:Cse}The self-evaporation coefficient $C_{se}^{\chi\to \xi}$.
It is easily seen from the plot that when $m_{\chi}>4\,{\rm GeV}$, the
effect is small enough to ignore it.}

\end{figure}

Likewise, evaporation can happen as a result of $\chi$ particle scattering
off $\xi$ particle in the Sun and vice versa. Hence we obtain coefficient of heterogeneous self-evaporation
is given by 
\begin{equation}
C_{se}^{\chi\to \xi}=\frac{\int4\pi r^{2}(dC_{se}^{\chi\to \xi}/dV)dr}{(\int4\pi r^{2}n_{\chi}(r)dr)(\int4\pi r^{2}n_{\xi}(r)dr)}\label{eq:Cse}
\end{equation}
where $dC_{se}^{\chi\to \xi}/dV$ is the heterogeneous self-evaporation in the Sun
within a given shell. It is a function of $m_\chi$, $m_\xi$ and $\sigma_{\chi\xi}$ and its analytical form is expressed in eq.~(\ref{eq:dCse/dV_general}).
If the
self-evaporation is due to either $\chi$ or $\xi$ itself, simply let 
 $m_{\chi}=m_{\xi}$ with additional factor of $1/2$ to avoid over counting.
The plot of $C_{se}^{\chi\to \xi}$ against
$m_{\chi}$ is given in figure~\ref{fig:Cse}.

In addition, if a DM particle in the halo transports enough kinetic
energy to the trapped DM particle in the Sun and leads to the trapped DM particle being ejected
to the interstellar space. This is called the (heterogeneous) self-ejection. 
Numerical calculations show the self-ejection effect is always small, compared to the self-capture one.
Thus, we can ignore
this effect safely from the calculation. Discussion on self-ejection
rate is presented in the appendix~\ref{sec:self-eject}.

\section{Numerical analysis: A model-independent treatment\label{sec:numerical_analysis}}

\begin{figure}
\begin{centering}
\includegraphics[width=0.45\textwidth]{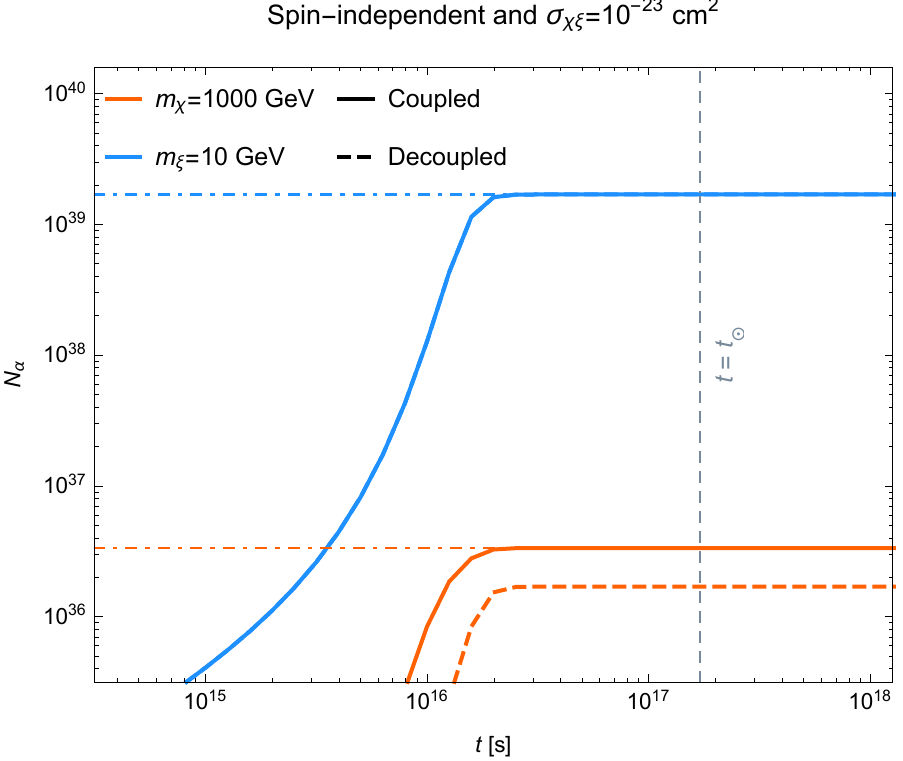}\quad\includegraphics[width=0.45\textwidth]{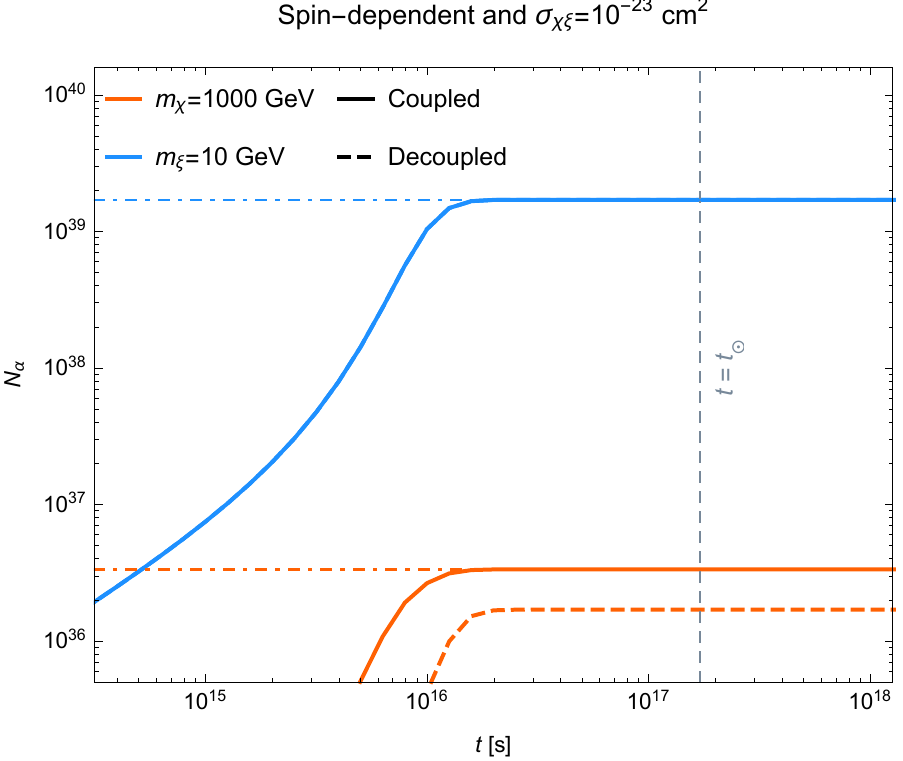}
\par\end{centering}
\begin{centering}
\includegraphics[width=0.45\textwidth]{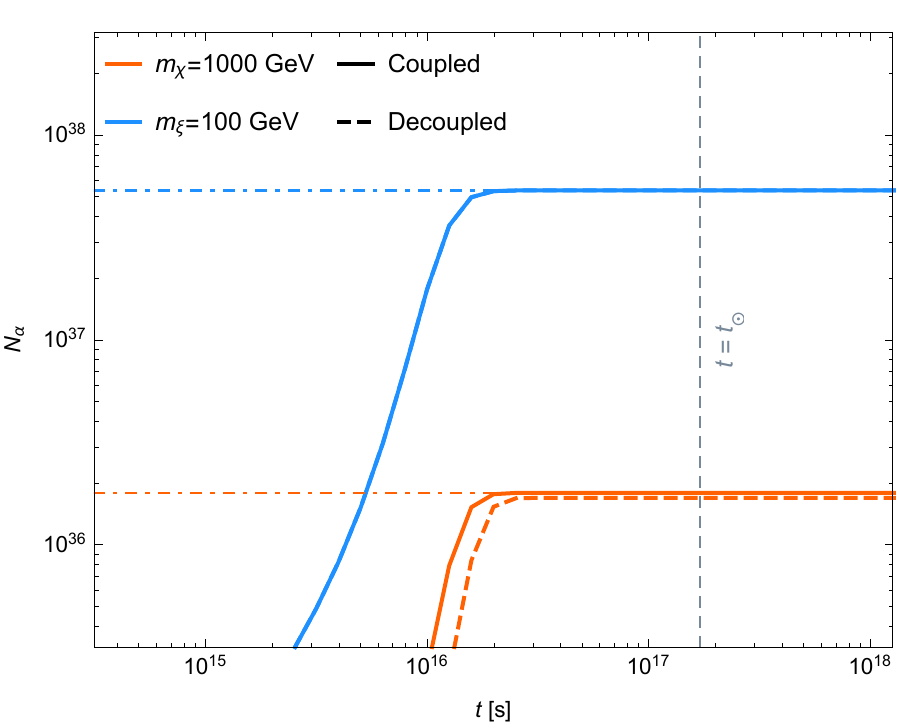}\quad\includegraphics[width=0.45\textwidth]{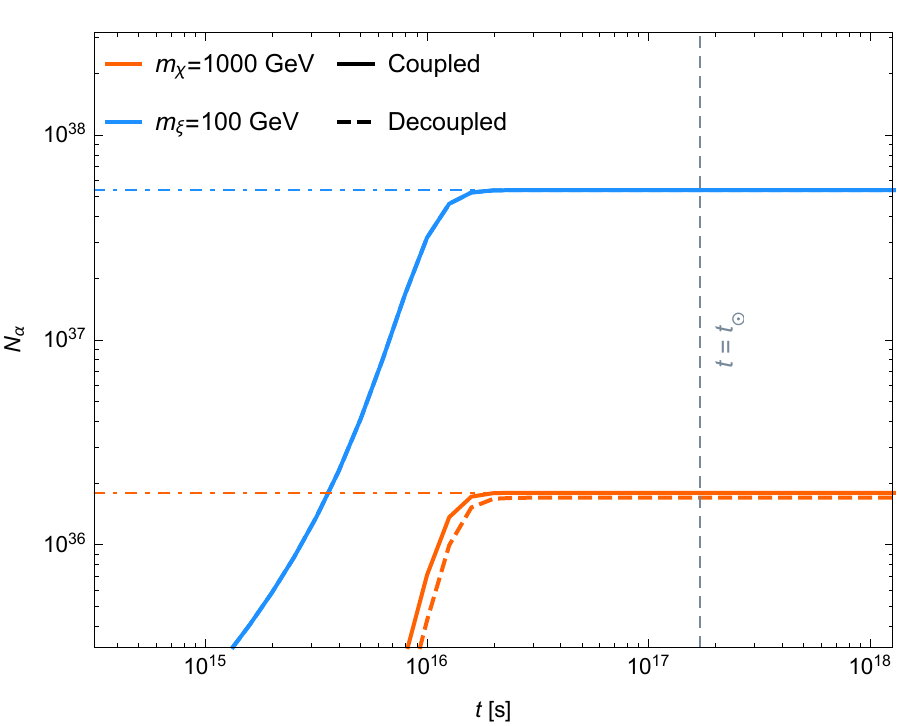}
\par\end{centering}
\caption{\label{fig:DM_evol}Evolution of DM numbers in the Sun with $r_{\rho}=1$.
We fixed $m_{\chi}=1000\,{\rm GeV}$ and from top to down $m_{\xi}=10\,{\rm GeV}$
and $100\,{\rm GeV}$ respectively. \emph{Left panels}: spin-independent.
\emph{Right panels:} spin-dependent. Gray dashed line indicates current
solar age $t=t_{\odot}\approx1.7\times10^{17}\,{\rm s}$. Dot-dashed
lines in each figure represent the $N_{\alpha}^{{\rm eq}}$ approximations
calculated from eqs.~(\ref{eq:N_eq_A}) and (\ref{eq:N_eq_B}). }
\end{figure}

\subsection{Number of dark matter particles in the Sun}

In the following analysis, when $r_{\rho}$ is assigned, the annihilation
cross section $\langle\sigma_{\alpha}v\rangle$ and $\rho_{\alpha}$
can be specified through eqs.~(\ref{eq:freeze_sigv}) and (\ref{eq:rho_x}).
Thus, thermal relic abundance $\Omega_{\rm DM}h^2\approx0.12$ and $\rho_{\rm DM}=0.3\,{\rm GeV}\,{\rm cm}^{-3}$
would be satisfied automatically. In the later numerical analysis, we have taken that $\sigma_{\alpha p}^{{\rm SI}}=10^{-46}\,{\rm cm}^{2}$
as a benchmark value for SI case. It is slightly smaller than the
most stringent value of LUX when the DM mass is roughly around $30\,{\rm GeV}$
\cite{Akerib:2016vxi}. For SD case, $\sigma_{\alpha p}^{{\rm SD}}=10^{-42}\,{\rm cm}^{2}$
that is chosen not to violate the results from Super-K \cite{Choi:2015ara}, PICO-60
\cite{Amole:2017dex} and IceCube \cite{Aartsen:2016zhm}. The self-scattering cross section
$\sigma_{\alpha}$ is indicated from eq.~(\ref{eq:sig_T}) and we set
$\sigma_{\chi\xi}=10^{-23}\,{\rm cm^2}$ and $10^{-24}\,{\rm cm^2}$ as the
benchmark values.
These two values are within 
$0.1\,{\rm cm^2\,g^{-1}}\leq \sigma_{{\rm DM}}/m_{{\rm DM}}\leq 10\,{\rm cm^2\,g^{-1}}$ in the whole interested mass range. The value of
$\sigma_{\chi\xi}$ tells us how $\chi$ and $\xi$ intertwine during the
evolution. The number of DM particles
in the Sun, $N_{\alpha}$, is plotted in figure~\ref{fig:DM_evol}
versus time $t$. We have fixed $m_{\chi}$ at $1000\,{\rm GeV}$ and
calculated with $m_{\xi}=100\,{\rm GeV}$ and $10\,{\rm GeV}$. The
case for $\sigma_{\chi\xi}=0$ is labeled as \emph{decoupled} for comparison. 

\begin{figure}
\begin{centering}
\includegraphics[width=0.45\textwidth]{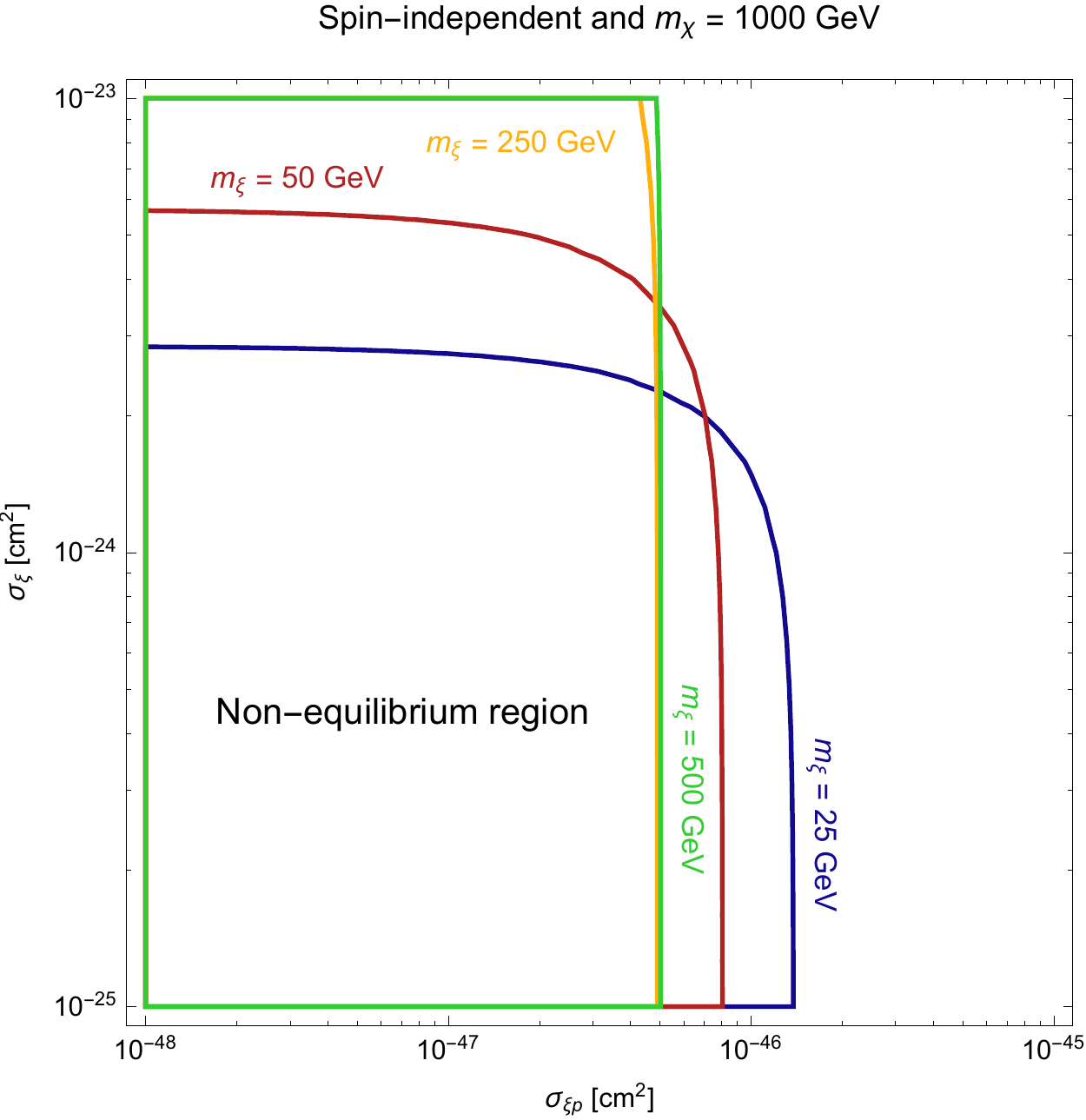}\quad\includegraphics[width=0.45\textwidth]{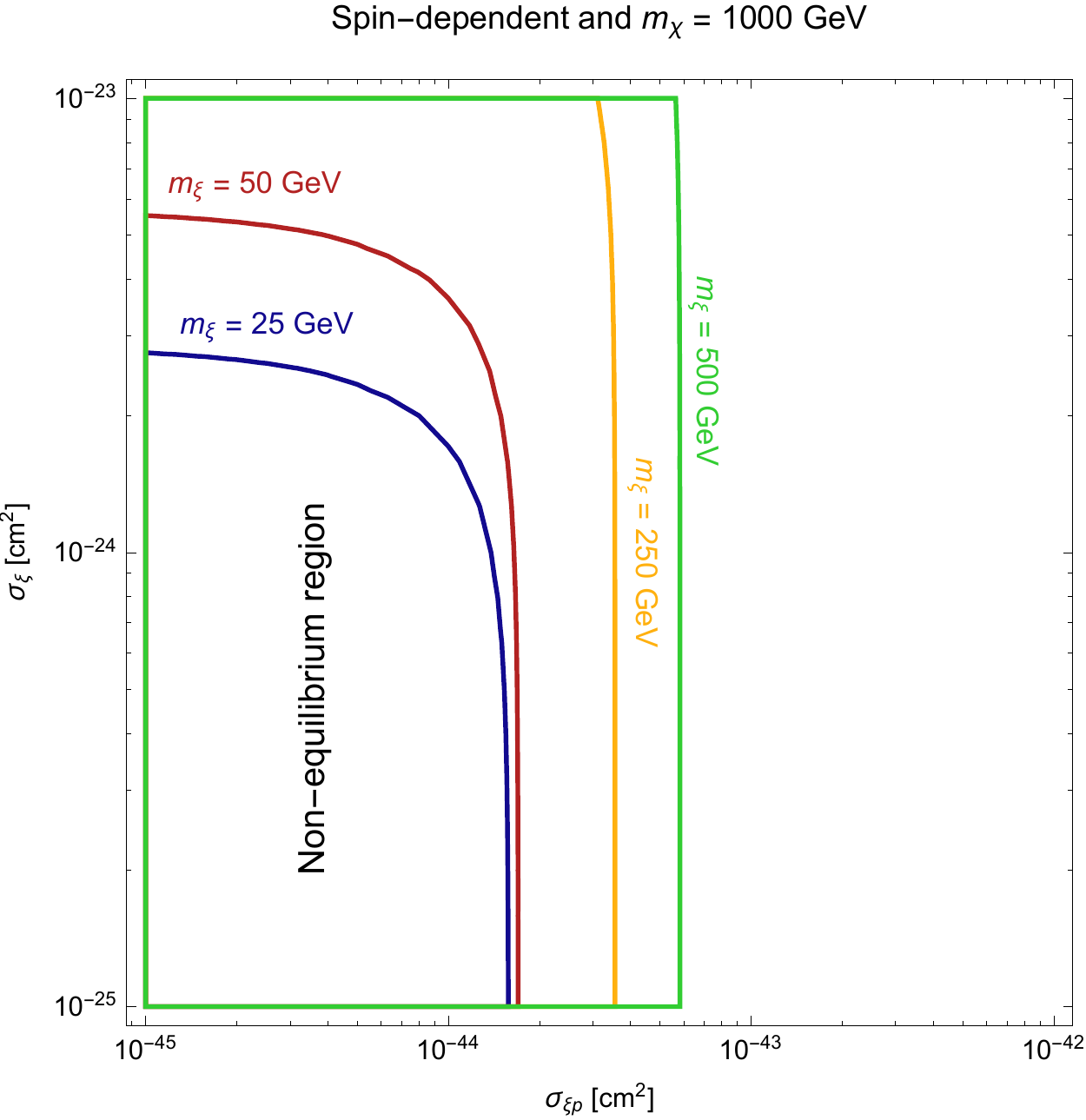}
\par\end{centering}
\caption{\label{fig:tau_eq}Equilibrium region for $m_{\xi}=25\,{\rm GeV}$(dark
blue), $50\,{\rm GeV}$ (dark red), $250\,{\rm GeV}$ (light orange)
and $500\,{\rm GeV}$ (lime green). \emph{Left}: SI case. \emph{Right}:
SD case. Region enclosed by each contour represents non-equilibrium,
$t_{\odot}/\tau<1$, at current epoch. It is assumed that $r_{\rho}=1$. In this choice of $m_\chi$ and $m_\xi$, the equilibrium timescale is always determined by $\xi$ solely. Thus, the effect of $\sigma_{\chi\xi}$ can be omitted. See main text for detail.}
\end{figure}

Number of DM in the Sun, $N_{{\rm \alpha}}$, is proportional to its
local number density $n_{{\rm \alpha}}=\rho_{\alpha}/m_{\alpha}$
when reaches equilibrium stage. Suppose $r_{\rho}=1$ and $m_{\chi}\gg m_{\xi}$.
We have $n_{\chi}\ll n_{\xi}$. Thus, $\chi$ affects little on the evolution of $\xi$. 
Hence, in the equilibrium stage, we could drop $C_{s}^{\xi\to \chi}N_{\chi}^{{\rm eq}}$
in eq.~(\ref{eq:B_no_eva}). In this way, simple expressions for
$N_{\chi}^{{\rm eq}}$ and $N_{\xi}^{{\rm eq}}$ can be given by
\begin{subequations}
\begin{align}
N_{\chi}^{{\rm eq}}&=\frac{C_{s}^{\chi}}{C_{a}^{\chi}}\left(\frac{1}{2}+\sqrt{\frac{1}{4}+R_{\chi}}\right)\label{eq:N_eq_A}\\
N_{\xi}^{{\rm eq}}&=\frac{C_{s}^{\xi}}{C_{a}^{\xi}}\left(\frac{1}{2}+\sqrt{\frac{1}{4}+R_{\xi}}\right)\label{eq:N_eq_B}
\end{align}
where
\begin{equation}
R_{\chi}=\frac{C_{a}^{\chi}(C_{c}^{\chi}+C_{s}^{\chi\to \xi}N_{\xi}^{{\rm eq}})}{(C_{s}^{\chi})^{2}}\quad{\rm and}\quad R_{\xi}=\frac{C_{a}^{\xi}C_{c}^{\xi}}{(C_{s}^{\xi})^{2}}\label{eq:R}
\end{equation}
\end{subequations}
are the correction factors due to the (heterogeneous) self-captures.
We have verified eqs.~(\ref{eq:N_eq_A}) and (\ref{eq:N_eq_B}) and
they agree with numerical solutions of eqs.~(\ref{eq:A_no_eva})
and (\ref{eq:B_no_eva}) very well after reaching the equilibrium state.
See dot-dashed lines in figure~\ref{fig:DM_evol}. When $n_{\xi}\gg n_{\chi}$,
$\xi$ evolves solely in the Sun. But $N_{\chi}^{{\rm eq}}$ is subject
to a correction that is proportional to $\sigma_{\chi\xi}N_{\xi}^{{\rm eq}}$.
In addition, we take eq.~(\ref{eq:sig_T}) as the benchmark
value of $\sigma_\alpha$. It results in a very strong self-interacting effect.
Therefore, the DM numbers in the equilibrium state, $N_{\alpha}^{\rm eq}$,
is mostly determined by the interactions in the DS. 
This agrees with the conclusion in ref.~\cite{Chen:2015poa} for 1DM
case.

On the other hand, equilibrium must achieves simultaneously for both
DM species. When $n_{\xi}\gg n_{\chi}$, the equilibrium timescale can be
determined by $\xi$ solely. It is given by $\tau_{\xi}=1/\sqrt{C_{c}^{\xi}C_{a}^{\xi}+(C_{s}^{\xi})^{2}/4}$.
Therefore, the role plays by $\sigma_{\chi\xi}$ that is
insignificant and can be omitted.
In figure~\ref{fig:tau_eq}, equilibrium region for a given $m_{\xi}$
is indicated by its corresponding color contour. The place enclosed
by the contour indicates $t_{\odot}/\tau<1$ as well as the non-equilibrium
region.

Note that when $m_{\chi}=m_{\xi}$, $C_{s}^{\chi\to \xi}=C_{s}^{\chi}$ and $C_{s}^{\xi\to \chi}=C_{s}^{\xi}$.
The evolution equations eqs.~(\ref{eq:A_evo}) and (\ref{eq:B_evo})
are degenerate. It can be considered as an 1DM scenario.

\subsection{Implication for the dark matter total annihilation rate in the Sun}

When an appreciated amount of DM particles accumulate in the solar
core, the total annihilation rate\footnote{The adjective \emph{total} does not imply summing over $\alpha$ but sum over all the DM number either from $\chi$ or $\xi$ in the Sun. 
	} as a result of these particles is given by
\begin{equation}
\Gamma_\alpha=\frac{1}{2}C_{a}^{\alpha}N_{\alpha}^{2}.
\end{equation}
for a given DM specie $\alpha$. If it is in the equilibrium state, we can apply eqs.~(\ref{eq:N_eq_A})
and (\ref{eq:N_eq_B}) and obtain
\begin{subequations}
\begin{align}
\Gamma_{\chi}^{{\rm eq}}&=\frac{1}{2}\frac{(C_{s}^{\chi})^{2}}{C_{a}^{\chi}}\left(\frac{1}{2}+\sqrt{\frac{1}{4}+R_{\chi}}\right)^{2},\label{eq:Gamma_1}\\
\Gamma_{\xi}^{{\rm eq}}&=\frac{1}{2}\frac{(C_{s}^{\xi})^{2}}{C_{a}^{\xi}}\left(\frac{1}{2}+\sqrt{\frac{1}{4}+R_{\xi}}\right)^{2},\label{eq:Gamma_2}
\end{align}
\end{subequations}
where $R_{\chi,\xi}$ is given in eq.~(\ref{eq:R}).
The above equations assume $\xi$ dominates the DM population over
$\chi$. Counter case is vice versa.

\begin{figure}
\begin{centering}
\includegraphics[width=0.45\textwidth]{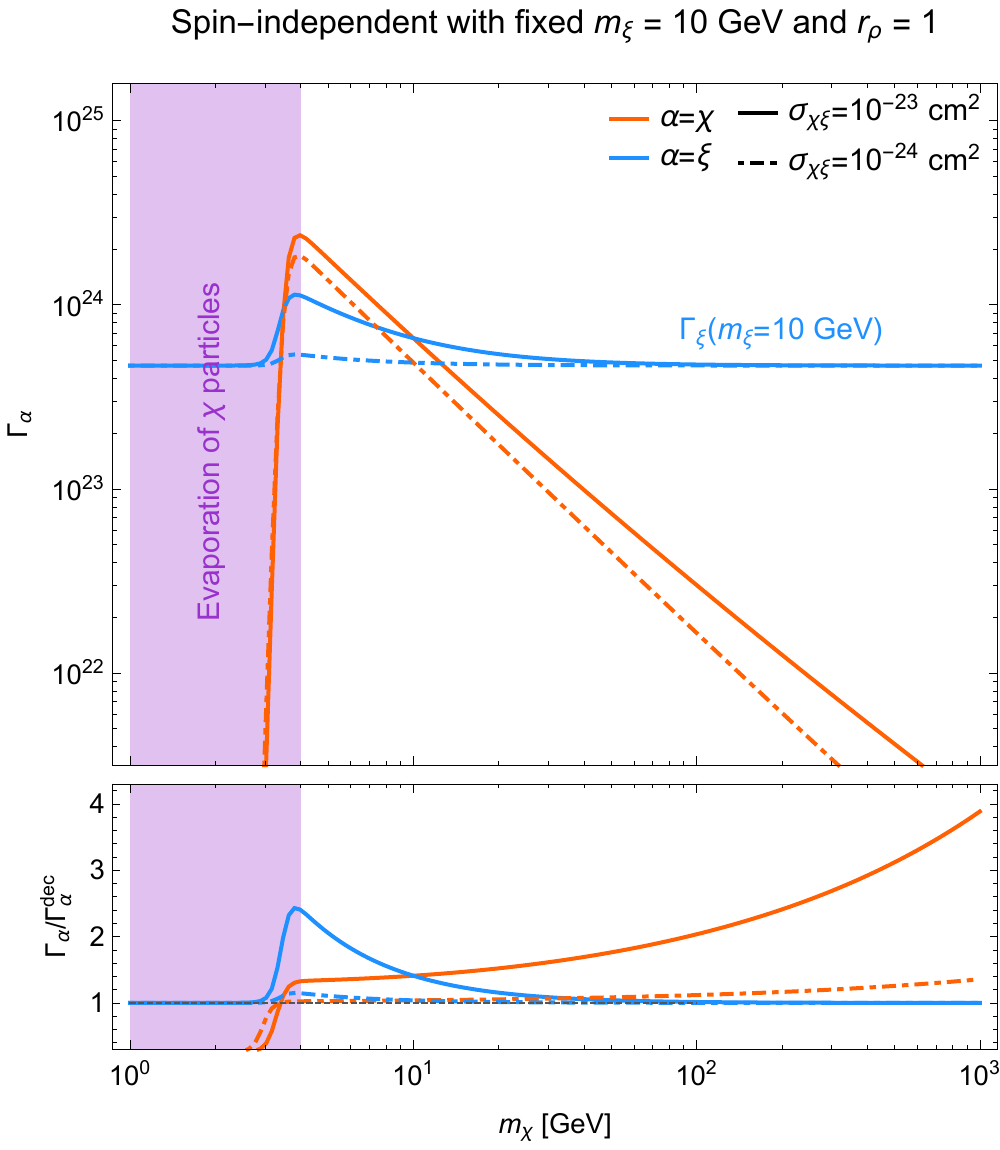}\quad\includegraphics[width=0.45\textwidth]{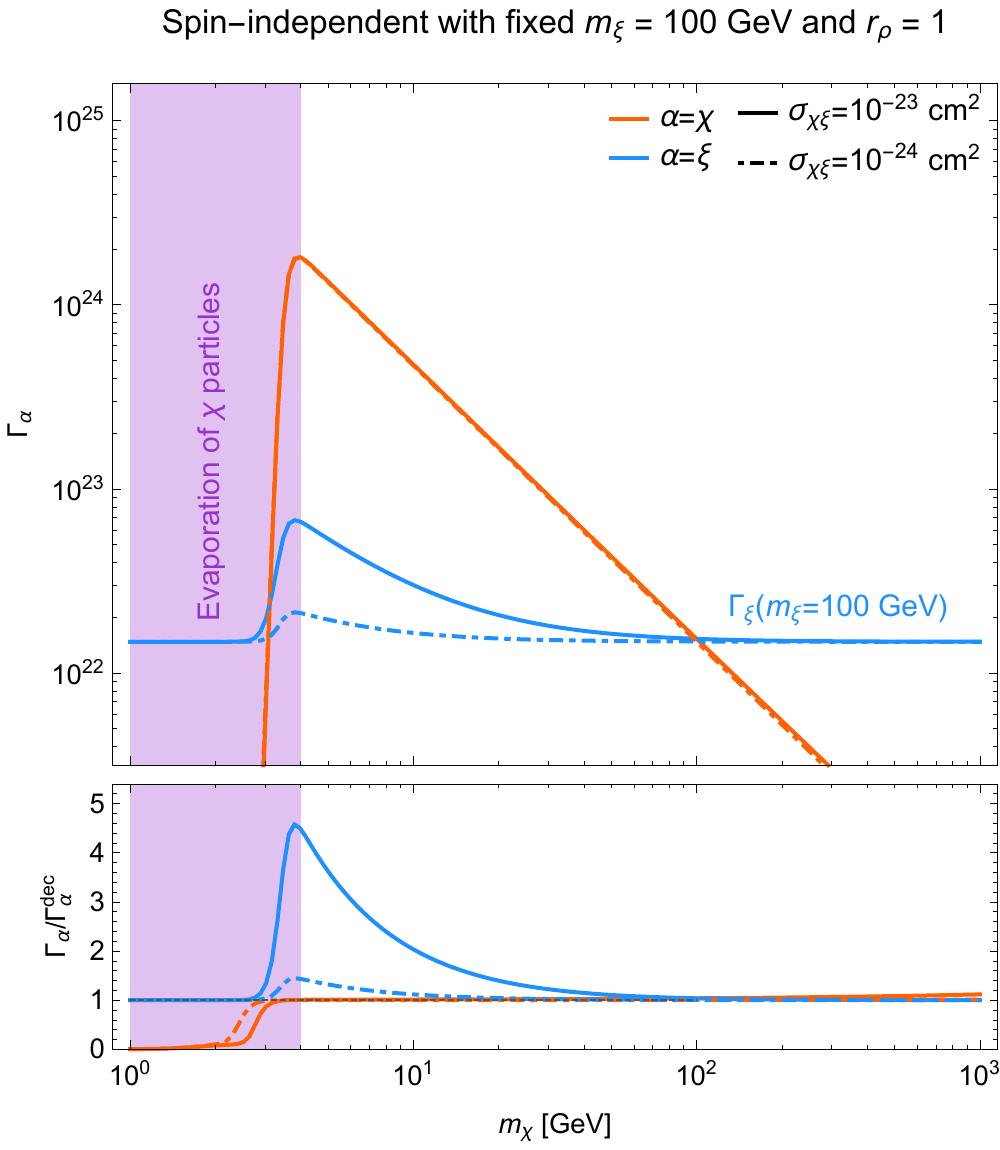}
\par\end{centering}
\caption{\label{fig:Gamma_r1}Total annihilation rate $\Gamma_{\alpha}$ with fixed
$m_{\xi}=10\,{\rm GeV}$ (left) and $100\,{\rm GeV}$ (right). Both
plots are calculated with $r_{\rho}=1$. Orange and blue lines are
for $\chi$ and $\xi$ particles respectively. Solid line indicates $\sigma_{\chi\xi}=10^{-23}\,{\rm cm}^{2}$
and dot-dashed $\sigma_{\chi\xi}=10^{-24}\,{\rm cm}^{2}$. $\Gamma_\alpha$ with
smaller $n_{\alpha}$ is subject to a larger correction from the other
specie. In the lower panel, the ratios between coupled and decoupled
are shown. Purple shaded region indicates the evaporation region of $\chi$.}
\end{figure}

For a more general discussion, unless specified, we will not assume
which specie is dominant over the other.
The plot of $\Gamma_\chi$ versus $m_{\chi}$ is shown in figure~\ref{fig:Gamma_r1}
with $r_{\rho}=1$. In this figure, we fixed $m_{\xi}=10\,{\rm GeV}$
and $100\,{\rm GeV}$ while $m_{\chi}$ runs from $1\,{\rm GeV}$ to
$1000\,{\rm GeV}$. 
In the above choice of parameters, both DM
species are all in equilibrium state today. As a consequence of large interactions
in the DS, the number of DM in the the equilibrium state is affected little from the
DM-nucleus interaction. Results from SI and SD cases are both similar.
Therefore, we focus on the SI case only in the following discussion.
In figure~\ref{fig:Gamma_r1}, the lower panel shows the ratio between coupled and decoupled cases.
Such ratio indicates how strong is the correction from $\sigma_{\chi\xi}$.

\begin{figure}
\begin{centering}
\includegraphics[width=0.45\textwidth]{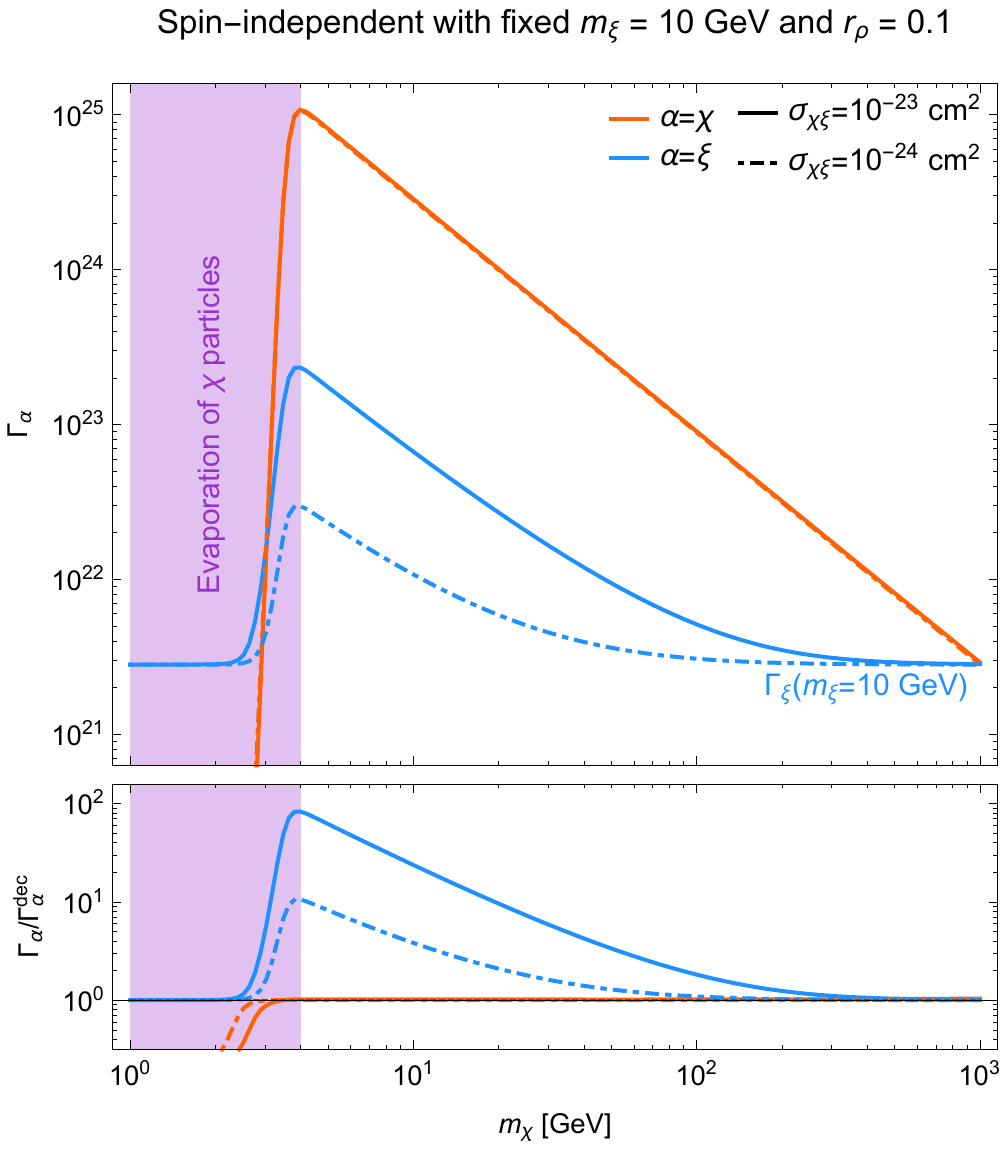}\quad\includegraphics[width=0.45\textwidth]{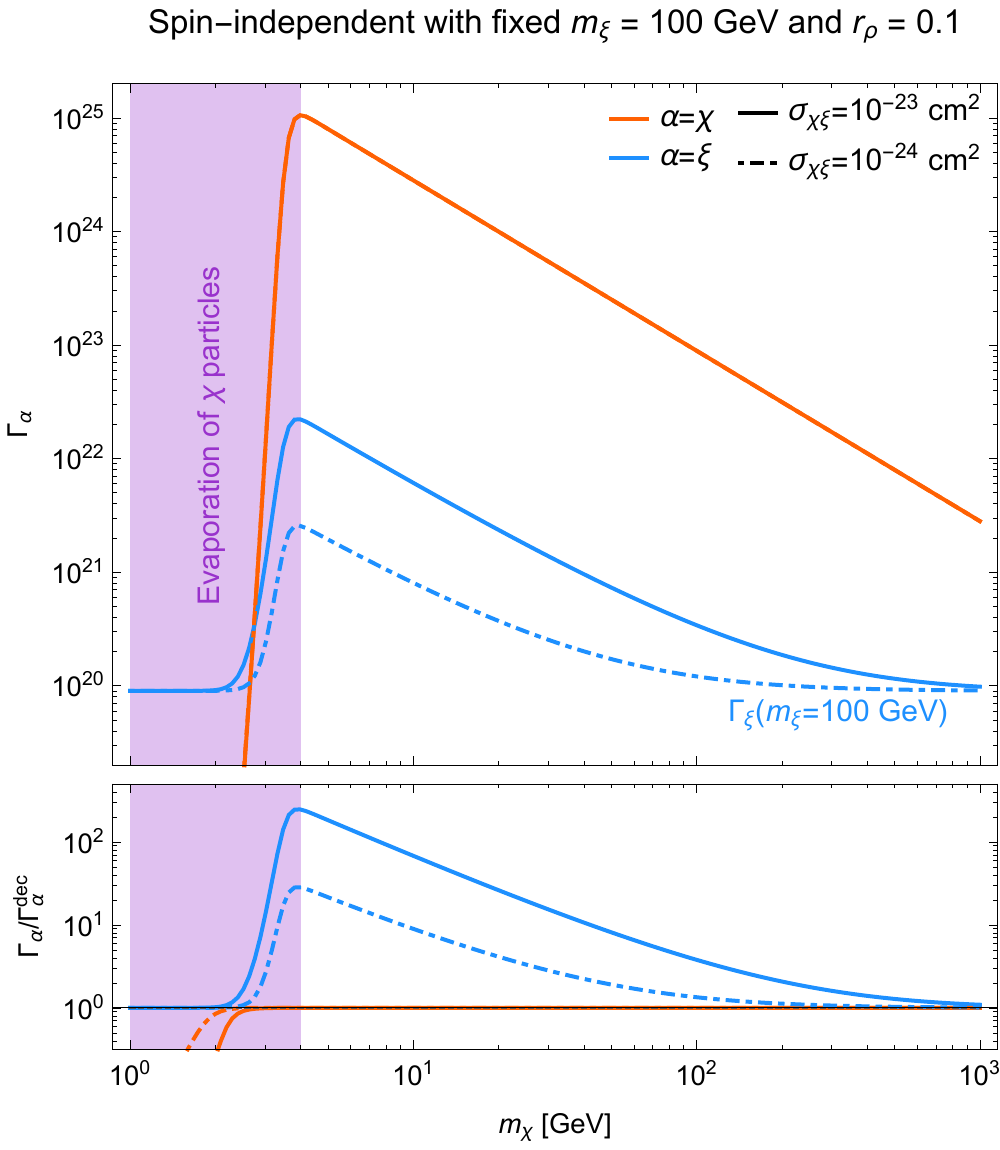}
\par\end{centering}
\begin{centering}
\includegraphics[width=0.45\textwidth]{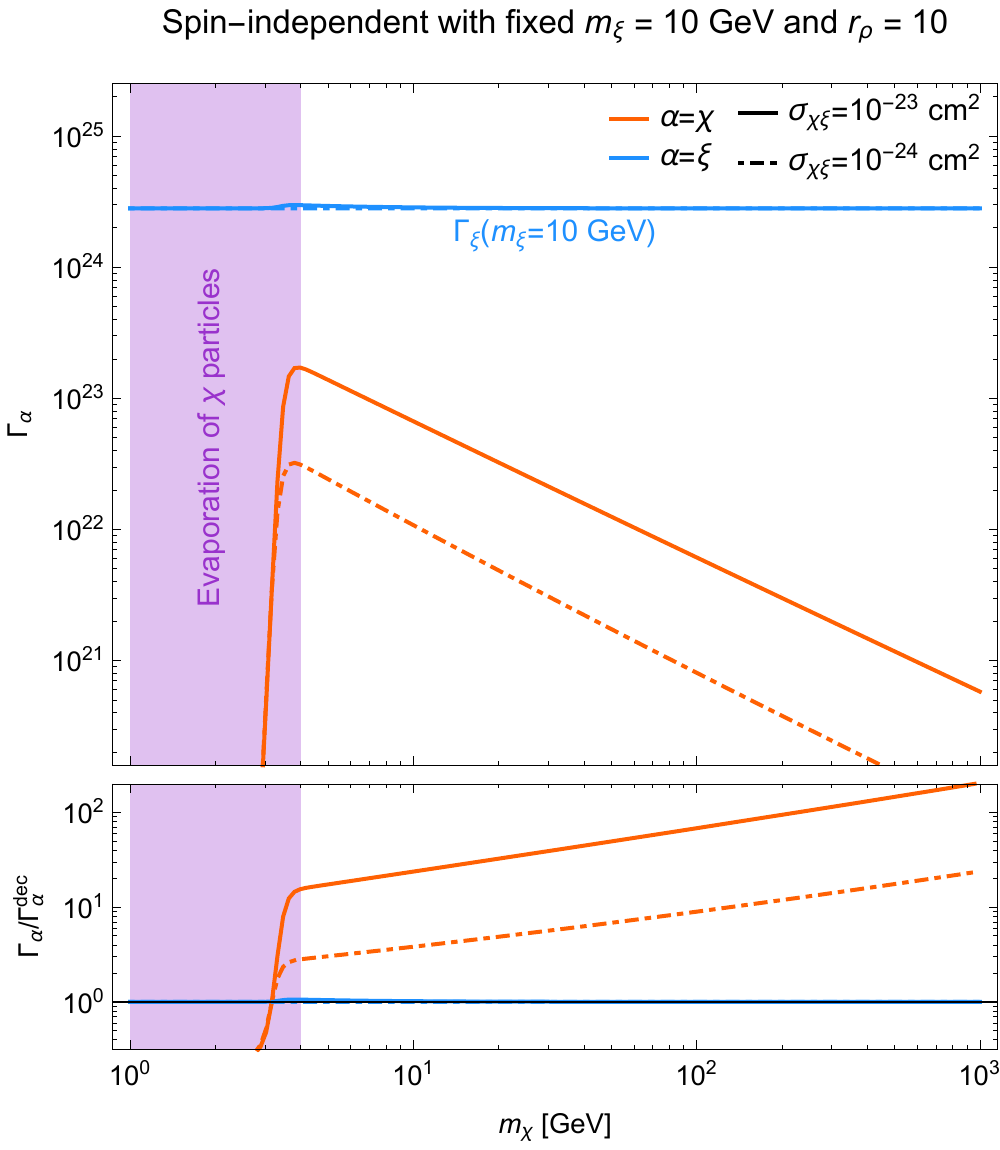}\quad\includegraphics[width=0.45\textwidth]{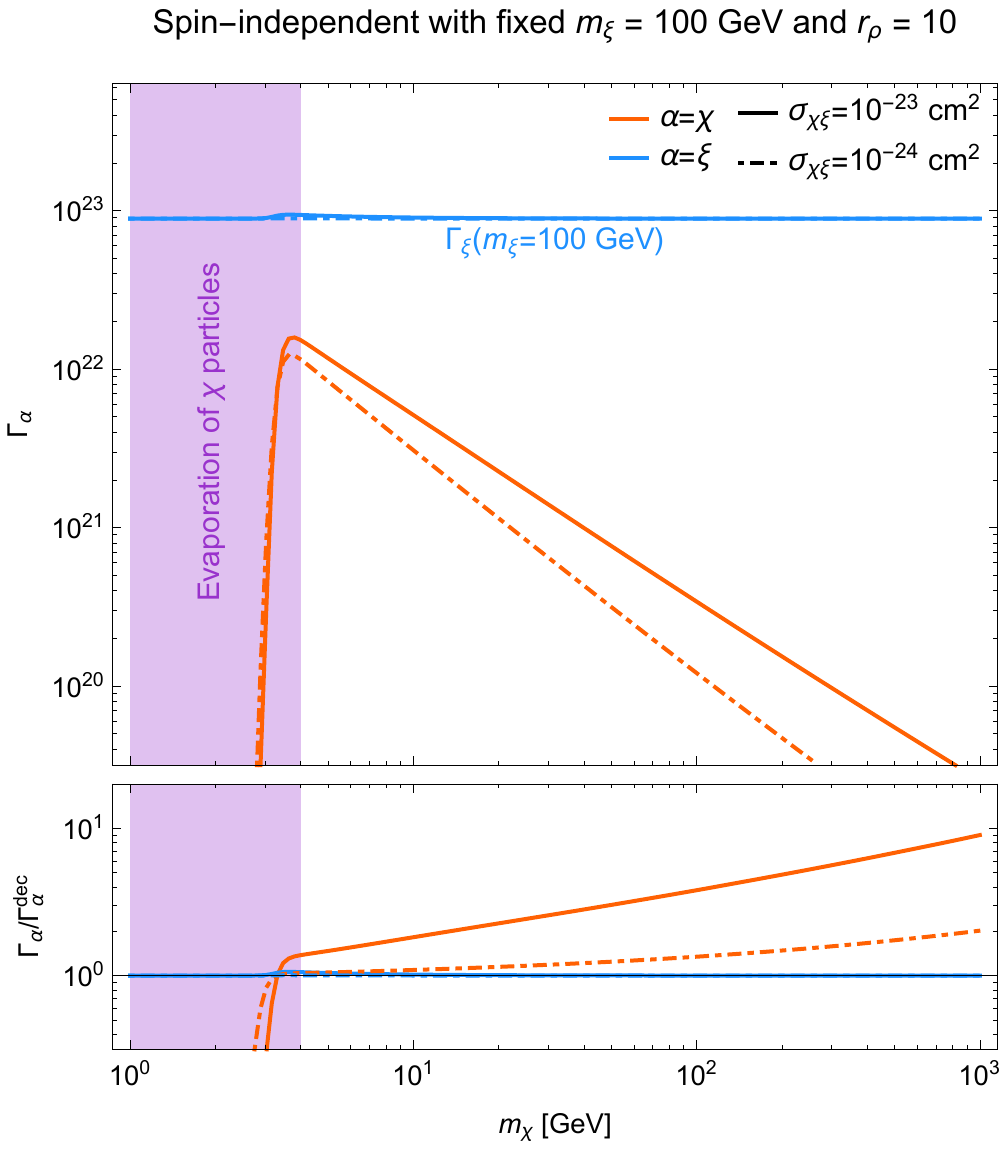}
\par\end{centering}
\caption{\label{fig:Gamma_asym_r}The same as figure~\ref{fig:Gamma_r1} but
$r_{\rho}=0.1$ (upper) and $10$ (lower).}
\end{figure}

On the left panel of figure~\ref{fig:Gamma_r1}, we fixed $m_\xi=10\,{\rm GeV}$ and $\Gamma_\xi$ is indicated by the blue line.
When $m_\chi>m_\xi$, it is true that $n_\chi<n_\xi$. Hence $\xi$ is the dominant
specie in the Sun and $\Gamma_\xi$ can be considered as independent of $\chi$ particles.
But $\Gamma_{\chi}$ is usually subject to a correction from $\xi$
when $m_\chi>m_\xi$. However, when $m_{\chi}$ is close to $m_{\xi}$, both
numbers $N_{\chi,\xi}$ are nearly equivalent. Mutual influence is strong
in this region. 
Not only $\Gamma_\chi$ is enhanced by $\xi$ particles, as well as
$\Gamma_\xi$ is increased by $\chi$ particles in the Sun. The ratio of correction
is shown in the lower panel. A quick drop of $\Gamma_{\chi}$ when $m_{\chi}\lesssim 4\,{\rm GeV}$
is due to the evaporation effect. The discussion is similar to the right
figure of figure~\ref{fig:Gamma_r1}, instead of raising $m_{\xi}$ to $100\,{\rm GeV}$.
Again, the correction from $\xi$ to $\Gamma_{\chi}$ is not significant when
$m_{\chi}>m_{\xi}$ here. Nonetheless, in the range $m_{\chi}$ is smaller than
$m_{\xi}$ ($n_{\chi}>n_{\xi}$), $\Gamma_{\xi}$ is subject to a correction
from $\chi$. Note that $m_{\chi}=100\,{\rm GeV}$ ($m_{\xi}=10\,{\rm GeV}$) of the left figure is
the same as the right figure of $m_{\chi}=10\,{\rm GeV}$ ($m_{\xi}=100\,{\rm GeV}$). It can be
realized from the symmetry of the evolution equations given in eqs.~(\ref{eq:A_no_eva}) and (\ref{eq:B_no_eva}).

The case for $r_{\rho}\neq1$ is shown in figure~\ref{fig:Gamma_asym_r}.
Parameters are the same as in the case of $r_{\rho}=1$ and the DM particles are
also in the equilibrium state. We know that $\Gamma_{\alpha}\propto C_{s}^{2}/C_{a}=n^{2}/\langle\sigma v\rangle$
and the ratio $\Gamma_{\chi}/\Gamma_{\xi}\propto (m_{\xi}/m_{\chi})^{2}/r_{\rho}^{3}$
in terms of eq.~(\ref{eq:r_sigv}). Hence we can deduce that $\Gamma_{\chi}/\Gamma_{\xi}\sim1$
when $m_{\chi}\approx32m_{\xi}$ for $r_{\rho}=0.1$. This statement is
partially correct since the exact $\Gamma_{\alpha}$ is subject to an extra
correction factor from $R_{\alpha}$ in eqs.~(\ref{eq:Gamma_1})
and (\ref{eq:Gamma_2}). We have numerically verified the correction factor
is roughly 3. Precisely speaking, when $m_{\chi}\approx100m_{\xi}$, $\Gamma_{\chi}/\Gamma_{\xi}\sim1$.
From the upper panel left in figure~\ref{fig:Gamma_asym_r}, it is
clearly seen that $\Gamma_{\chi}/\Gamma_{\xi}\sim1$ when $m_{\chi}\sim1000\,{\rm GeV}$.
This argument agrees with our numerical result well. Similarly, it applies to the case of $m_{\xi}=100\,{\rm GeV}$.
In this case, $\Gamma_{\chi}/\Gamma_{\xi}\sim1$ when $m_{\chi}\approx10^{4}\,{\rm GeV}$.
For $r_{\rho}=10$, we can use the approximation above and obtain that
$\Gamma_{\chi}/\Gamma_{\xi}\sim1$ when $m_{\chi}\approx0.01m_{\xi}$. Therefore, $m_{\chi}=0.1(1)\,{\rm GeV}$ when $m_{\xi}=10(100)\,{\rm GeV}$
that we would have $\Gamma_{\chi}/\Gamma_{\xi}\sim1$. However, one should bear in mind that for $m_{\chi} \lesssim4$~GeV all $\chi$ particles have evaporated already. 

\section{Summary\label{sec:Summary}}

In this paper, we address the issue of 2DM evolution in the Sun. We
consider a scenario, where the heterogeneous $\chi\xi$ self-scattering happens.
Such interaction weaves the evolution processes for both DM species that was
assumed to evolve independently.
We found that when one DM specie is sub-dominant, its number of particles in the Sun is
subject to a correction from the dominant specie. This correction always enhances the number
of DM particles being captured. When the masses of the two DM species are close, the enhancement is mutual and has the largest impact.
However, the sub-dominant specie in general has smaller total annihilation
rate, the effect of heterogeneous self-capture would be tiny to the detection unless its annihilation final state is distinct
from the dominant one.

Though the heterogeneous self-interaction causes extra self-evaporation and
self-ejection, we have demonstrated that these negative effects are either small or
it must happens when the DM mass is sufficient light, $m_{\alpha}^{\rm ev}\lesssim 4\,{\rm GeV}$.
Therefore, in most of the interested mass range that relates to our study, they can be safely
ignored.

To summarize, we would like to point out that the heterogeneous self-interaction is a 
natural consequence of any 2DM or \emph{n}DM models. This effect will eventually reflect in the DM 
annihilation rates. Potentially, if the DM annihilates to the SM particles in the final state,
such signal could be detected in the terrestrial detectors. Therefore, the strength of the
heterogeneous self-interaction could be probed. Moreover, any sign of such interaction
could be considered as a possible existence of DM beyond one-component.

\acknowledgments
C.~S. Chen (TKU) and Y.~H. Lin (NCKU) are supported by the Ministry
of Science and Technology, Taiwan under Grant No.~104-2112-M-032-009-MY3 
and 106-2811-M-006-041 respectively.

\appendix

\section{Derivations of the 2DM heterogeneous self-scattering rates}
\subsection{The self-capture rate\label{sec:self_cap}}

To the capture rate of different classes of particles has been fully
discussed in refs.~\cite{Gould:1987ju,Gould:1987ir,Zentner:2009is}.
In this appendix, we only present the mathematical key point to derive
the heterogeneous self-capture rate.

Following earlier works \cite{Gould:1987ir,Zentner:2009is}, the problem
begins by considering capture in a spherical shell of material (solar
interior) on which capture is happening of radius $r$ and local escape
velocity $v_{{\rm esc}}(r)$. Now at an imaginary surface bounding
a region of radius $R$, which the solar gravity is negligible at
$R$. The DM flux goes inward across the surface is \cite{Press:1985ug}
\begin{equation}
\pi R^{2}f(u)udu\frac{dJ^{2}}{R^{2}u^{2}}\label{eq:DM_flux_R}
\end{equation}
where $f(u)$ is the DM velocity distribution at infinity, $J=Ru\sin\theta$
the angular momentum per unit mass and $\theta$ the angle relative
to the radial direction. Taking $\Omega(w)$ is the rate at which
a DM particle enters the shell $r$ with velocity $w=\sqrt{v_{{\rm esc}}^{2}(r)+u^{2}}$
and scatters to velocity less than $v_{{\rm esc}}(r)$. The probability
of such a DM to be captured is \cite{Zentner:2009is}
\begin{equation}
dP=\frac{\Omega(w)}{w}\frac{2dr}{\sqrt{1-J^{2}/(rw)^{2}}}\Theta(rw-J)\label{eq:cap_p}
\end{equation}
where $\Theta$ is the Heaviside step function. The differential rate
of capture can be easily obtained by multiplying eqs.~(\ref{eq:DM_flux_R})
and (\ref{eq:cap_p}) then integrate over all angular momentum $J^{2}$.
Replacing $dV=4\pi r^{2}dr$ we have,
\begin{equation}
\frac{dC}{dVdu}=\frac{f(u)}{u}w\Omega(w).
\end{equation}
Thus, the total DM capture rate per unit shell volume is given by
\begin{equation}
\frac{dC}{dV}=\int\frac{f(u)}{u}w\Omega(w)du.\label{eq:dC/dV}
\end{equation}
In the above equation, $w$ depends on $u$ explicitly. The remaining
task is to determine $\Omega(w)$.

The scattering in the shell is simply $n\sigma w$, with the the scattering
cross section $\sigma$ and the target number density $n$. Practically
we assume nearly isotropic and velocity-independent $\sigma$. The
incoming particle with $m_{\chi}$ and scatters off bounded particle
with $m_{\xi}$. In order to be captured, $\chi$ particle must loses a fractional
of kinetic energy over the interval
\begin{equation}
\frac{u^{2}}{w^{2}}\leq\frac{\Delta E}{E}\leq\frac{\mu}{\mu_{+}^{2}}\label{eq:posi_cond}
\end{equation}
where $\mu$ and $\mu_{\pm}$ are expressed as 
\begin{gather}
\mu=\frac{m_{\chi}}{m_{\xi}},\quad\mu_{\pm}=\frac{\mu\pm1}{2},\label{eq:mu}
\end{gather}
and
$\eta^{2}=3(v_{\odot}/\bar{v})^{2}/2$,
$v_{\odot}=220\,{\rm km}\,{\rm s}^{-1}$ the solar moving velocity
and $\bar{v}=270\,{\rm km}\,{\rm s}^{-1}$ the DM velocity dispersion.

Therefore, the capture probability in each scattering is
\begin{equation}
p_{{\rm cap}}=\frac{\mu_{+}^{2}}{\mu}\left(\frac{\mu}{\mu_{+}^{2}}-\frac{u^{2}}{w^{2}}\right)\Theta\left(\frac{\mu}{\mu_{+}^{2}}-\frac{u^{2}}{w^{2}}\right).\label{eq:p_cap}
\end{equation}
The rate of capture is simply the scattering rate $n_{\xi}\sigma w$
times the capture probability $p_{{\rm cap}}$. Hence, 
\begin{equation}
\Omega(w)=n_{\xi}\sigma wp_{{\rm cap}}=\frac{\sigma n_{\xi}v_{{\rm esc}}^{2}(r)}{w}\left[1-\frac{u^{2}}{v_{{\rm esc}}^{2}(r)}\frac{\mu_{-}^{2}}{\mu}\right]\Theta\left(1-\frac{u^{2}}{v_{{\rm esc}}^{2}(r)}\frac{\mu_{-}^{2}}{\mu}\right).\label{eq:Omega}
\end{equation}
Combining eqs.~(\ref{eq:dC/dV}) and (\ref{eq:Omega}) we have
\begin{align}
\frac{dC_{s}^{\chi\to \xi}}{dV} & =\int\sigma n_{\xi}(r)v_{{\rm esc}}^{2}(r)\frac{f(u)}{u}\left(1-\frac{u^{2}}{v_{{\rm esc}}^{2}(r)}\frac{\mu_{-}^{2}}{\mu}\right)\Theta(v_{{\rm esc}}^{2}(r)-\mu u^{2})\label{eq:dC/dV2}
\end{align}
With respect to the solar moving frame, we can expressed $f(u)$ as
\begin{equation}
f(u)=\frac{4}{\sqrt{\pi}}n_{\chi}x^{2}e^{-x^{2}}e^{-\eta^{2}}\frac{\sinh(2x\eta)}{2x\eta}\label{eq:v_dist}
\end{equation}
in terms of the dimensionless variables $x^{2}=3(u/\bar{v})^{2}/2$
and $\eta^{2}=3(v_{\odot}/\bar{v})^{2}/2$. Integrating eq.~(\ref{eq:dC/dV2})
over $u$, we have
\begin{align}
\frac{dC_{s}^{\chi\to \xi}}{dV}=&  \sqrt{\frac{3}{2}}n_{\chi}n_{\xi}(r)\sigma_{\chi\xi}\frac{v_{{\rm esc}}^{2}(r)}{2\eta\bar{v}Y^{2}}\left\{ \left(Y_{+}Y_{-}-\frac{1}{2}\right)[X(-\eta,\eta)-X(Y_{-},Y_{+})]\right.\nonumber \\
 & \left.+\frac{1}{2}Y_{+}e^{-Y_{-}^{2}}-\frac{1}{2}Y_{-}e^{-Y_{+}^{2}}-\eta e^{-\eta^{2}}\right\} \label{eq:dCs/dV_general}
\end{align}
where we have replaced $\sigma$ by $\sigma_{\chi\xi}$ to indicate the heterogeneous
self-scattering cross section and
\begin{subequations}
\begin{gather}
Y^{2}=\frac{3}{2}\frac{v_{{\rm esc}}^{2}(r)}{\bar{v}^{2}}\frac{\mu}{\mu_{-}^{2}},\quad Y_{\pm}=Y\pm\eta,\\
X(a,b)\equiv\int_{a}^{b}e^{-y^{2}}dy=\frac{\sqrt{\pi}}{2}[{\rm erf}(b)-{\rm erf}(a)].
\end{gather}
\end{subequations}
Thus, the coefficient of heterogeneous self-capture rate
is evaluated as
\begin{equation}
C_{s}^{\chi\to \xi}=\frac{\int4\pi r^{2}(dC_{s}^{\chi\to \xi}/dV)dr}{\int4\pi r^{2}n_{\xi}(r)dr}\label{eq:Cs_gen}
\end{equation}
where $n_{\xi}(r)$ is the number distribution of $\xi$ particles in
the Sun. The case for halo $\xi$ particle scatters with solar trapped $\chi$ particle is
essentially identical.

\subsection{The self-evaporation rate\label{sec:self_evap_2DM}}

Self-evaporation happens when two DM particles collide, one gets velocity
larger than the escape velocity $v_{{\rm esc}}$. Such calculation
is similar to the evaporation between DM and nucleus presented in
ref.~\cite{Gould:1987ju}. Here we show the key to obtain the heterogeneous self-evaporation
rate.

To scatter a $\xi$ particle from velocity $w$ to $v_{{\rm esc}}>w$
by $\chi$ particle, the rate is
\begin{align}
\Omega_{se}(w)= & \sqrt{\frac{2}{\pi}}\frac{T_{\chi}}{m_{\xi}}\frac{1}{\mu^{2}}\frac{\sigma_{\chi\xi}n_{\chi}(r)}{w}\bigg\{
\mu(\alpha_{+}e^{-\alpha_{-}^{2}}-\alpha_{-}e^{-\alpha_{+}^{2}})\nonumber \\
 & +2\mu_{+}^{2}X(\beta_{-},\beta_{+})\exp\left[-\frac{m_{\xi}}{2T_{\chi}}(v_{{\rm esc}}^{2}(r)-w^{2})\right]\nonumber\\
 & +(\mu-2\mu\alpha_{+}\alpha_{-}-2\mu_{+}\mu_{-})X(\alpha_{-},\alpha_{+})\bigg\}
\end{align}
where 
\begin{subequations}
	\begin{align}
	\alpha_{\pm}=&\sqrt{\frac{m_{\chi}}{T_{\chi}}}(\mu_{+}v_{{\rm esc}}(r)\pm\mu_{-}w),\label{eq:apm}\\
	\beta_{\pm}=&\sqrt{\frac{m_{\xi}}{T_{\xi}}}(\mu_{-}v_{{\rm esc}}(r)\pm\mu_{+}w).\label{eq:bpm}
	\end{align}
\end{subequations}
Assuming $\xi$ particles are in a truncated Maxwell-Boltzmann distribution
with a cutoff velocity $w_{c}$,
\begin{equation}
f(w)dw=\frac{4}{\sqrt{\pi}}\left(\frac{m_{\xi}}{2T_{\xi}}\right)^{3/2}n_{\xi}(r)w^{2}e^{-m_{\xi}w^{2}/(2T_{\xi})}\Theta(w_{c}-w)dw.
\end{equation}
Thus,
\begin{equation}
\frac{dC_{se}^{\chi\to \xi}}{dV}=\int f(w)\Omega_{se}(w)dw\label{eq:Cse_2DM}
\end{equation}
In order to evaluate eq.~(\ref{eq:Cse_2DM}), we assumed the that
$T_{\chi}=T_{\xi}=T$ and $w_{c}=v_{{\rm esc}}$.\footnote{DM temperature could depend on its mass and in general $T_{\chi}/T_{\xi}\neq1$.
However, the deviation from unity is small \cite{Garani:2017jcj}.}
Therefore,
\begin{align}
\frac{dC_{se}^{\chi\to \xi}}{dV}= & \frac{2}{\pi}\sqrt{\frac{2T}{m_{\xi}}}n_{\xi}(r)n_{\chi}(r)\sigma_{\chi\xi}\left[e^{-E_{e}/T}\left(-\beta_{+}\beta_{-}-\frac{1}{2\mu}\right)X(\beta_{-},\beta_{+})\right.\nonumber \\
 & \left.+e^{-E_{e}/T}\left(\alpha_{+}\alpha_{-}-\frac{1}{2\mu}\right)X(\alpha_{-},\alpha_{+})+e^{-(E_{e}/T+\alpha_{+}^{2})}\sqrt{\frac{m_{\chi}}{2T}}v_{{\rm esc}}(r)\right] \label{eq:dCse/dV_general}
\end{align}
where $E_{e}=m_{\xi}v_{{\rm esc}}^{2}(r)/2$. 
Therefore, we have the coefficient of the heterogeneous self-evaporation
rate,
\begin{equation}
C_{se}^{\chi\to \xi}  =\frac{\int4\pi r^{2}(dC_{se}^{\chi\to \xi}/dV)dr}{(\int4\pi r^{2}n_{\chi}(r)dr)(\int4\pi r^{2}n_{\xi}(r)dr)}.
\end{equation}
When $m_{\chi}=m_{\xi}$, it reduces to the 1DM case and a symmetric factor
$1/2$ should be introduced to avoid over counting.

\subsection{The self-ejection rate\label{sec:self-eject}}

Once the incoming $\chi$ particle loses a fraction of energy $\Delta E/E>v_{{\rm esc}}^{2}(r)/w^{2}$
to a trapped particle $\xi$. The $\xi$ particle will be ejected from
the Sun. Following the derivation in the appendix~\ref{sec:self_cap} but replacing $p_{{\rm cap}}$ by the ejection probability \cite{Zentner:2009is}
\begin{equation}
p_{{\rm ejec}}=\frac{\mu_{+}^{2}}{\mu}\left(\frac{\mu}{\mu_{+}^{2}}-\frac{v_{{\rm esc}}^{2}(r)}{w^{2}}\right)\Theta\left(\frac{\mu}{\mu_{+}^{2}}-\frac{v_{{\rm esc}}^{2}(r)}{w^{2}}\right).
\end{equation}
Thus, the rate of ejection,
\begin{equation}
\Omega_{ej}(w)=\frac{n_{\xi}(r)\sigma_{\chi\xi}}{w}\left(u^{2}-\frac{\mu_{-}^{2}}{\mu}v_{{\rm esc}}^{2}(r)\right)\Theta\left(u^{2}-\frac{\mu_{-}^{2}}{\mu}v_{{\rm esc}}^{2}(r)\right).
\end{equation}
Integrating over the $\chi$ number distribution in the halo $f(u)$
given in eq.~(\ref{eq:v_dist}) we have
\begin{equation}
\frac{dC_{ej}^{\chi\to \xi}}{dV}=\int\sigma_{\chi\xi} n_{\xi}(r)\frac{f(u)}{u}\left(u^{2}-\frac{\mu_{-}^{2}}{\mu}v_{{\rm esc}}^{2}(r)\right)\Theta\left(u^{2}-\frac{\mu_{-}^{2}}{\mu}v_{{\rm esc}}^{2}(r)\right)du.
\end{equation}
By changing of variable we get
\[
\frac{dC_{ej}^{\chi\to \xi}}{dV}=\int\frac{\sigma_{\chi\xi} n_{\xi}(r)}{K^{2}}\frac{\mu_{-}^{2}}{\mu}v_{{\rm esc}}^{2}(r)\frac{f(x)}{x}(x^{2}-K^{2})\Theta(x-K)dx
\]
where
\[
K^{2}=\frac{3}{2}\frac{v_{{\rm esc}}^{2}(r)}{\bar{v}^{2}}\frac{\mu_{-}^{2}}{\mu}\quad{\rm and}\quad K_{\pm}=K\pm\eta.
\]
Therefore,
\begin{align}
\frac{dC_{ej}^{\chi\to \xi}}{dV}=&\frac{4}{\sqrt{\pi}}\sigma_{\chi\xi} n_{\chi}n_{\xi}(r)\frac{\bar{v}}{3\eta}\bigg[e^{-K_{+}^{2}}(e^{4K\eta}K_{+}-K_{-})\nonumber\\
&-\frac{1}{2}\left(K_{+}K_{-}-\frac{1}{2}\right)X(K_{-},K_{+})\bigg].
\end{align}
Our final result of the heterogeneous self-ejection rate is evaluated as
\begin{equation}
C_{ej}^{\chi\to \xi}=\frac{\int4\pi r^{2}(C_{ej}^{\chi\to \xi}/dV)dr}{\int4\pi r^{2}n_{\xi}(r)dr}.
\end{equation}
However, due to the large escape velocity in the Sun, such self-ejection effect
is always insignificant comparing to other effects.
Thus, we can safely ignore this correction in the DM evolution.

\end{document}